\newcommand{\blind}{1}

\documentclass[12pt]{article}

\usepackage{amsmath,epsfig,amssymb,bm,amsfonts,amsthm,epsfig,verbatim, enumitem}
\usepackage[comma,authoryear]{natbib}
\usepackage{color,graphicx,lscape,longtable,mathabx,natbib, float, multirow}
\usepackage{hyperref,subfigure, multirow, setspace,  booktabs}
\usepackage{algorithm}
\usepackage{algorithmicx}
\usepackage{algpseudocode}
\graphicspath{{figures/}}

\renewcommand{\baselinestretch} {1.6}
\makeatletter 
\def\singlespace{\def\baselinestretch{1}\@normalsize}

\date{}
\newcommand{\bit}{\begin{itemize}}
\newcommand{\eit}{\end{itemize}}

\newcommand{\normmm}[1]{{\left\vert\kern-0.25ex\left\vert\kern-0.25ex\left\vert #1
    \right\vert\kern-0.25ex\right\vert\kern-0.25ex\right\vert}}

\@addtoreset{equation}{section}
\newtheorem{Theorem}{Theorem}[]

\newcommand{\var}{\mbox{VAR}}

\renewcommand{\hat}{\widehat}
\def\bse{\begin{eqnarray*}}
\def\ese{\end{eqnarray*}}
\def\be{\begin{eqnarray}}
\def\ee{\end{eqnarray}}
\def\bsq{\begin{equation*}}
\def\esq{\end{equation*}}
\def\bq{\begin{equation}}
\def\eq{\end{equation}}

\def\var{\hbox{var}}

\def\wh{\widehat}

\def\ba{\boldsymbol\alpha}
\def\bLambda{\boldsymbol\Lambda}

\def\bvarepsilon{\boldsymbol\varepsilon}

\def\bb{{\boldsymbol\beta}}

\def\bxi{\boldsymbol\xi}

\def\bg{\boldsymbol\gamma}

\def\Delta{\boldsymbol\Delta}

\def\btheta{\boldsymbol\theta}
\def\G{\boldsymbol G}

\newcommand{\nvs}{\vspace{-0.3in}}

\def\A{{\bf A}}
\def\U{{\bf U}}
\def\V{{\bf V}}

\def\a{{\bf a}}

\def\B{{\bf B}}

\def\D{{\bf D}}
\def\V{{\bf V}}
\def\K{{\bf K}}

\def\r{{\bf r}}
\def\f{{\bf f}}
\def\h{{\bf h}}

\def\I{{\bf I}}

\def\F{{\bf F}}
\def\H{{\bf H}}
\def\m{{\bf m}}

\def\K{{\bf K}}

\def\U{{\bf U}}
\def\S{{\bf S}}

\def\u{{\bf u}}
\def\m{{\bf m}}

\def\X{{\bf X}}
\def\S{{\bf S}}
\def\x{{\bf x}}
\def\I{{\bf I}}

\def\Y{{\bf Y}}

\def\y{{\bf y}}
\def\0{{\bf 0}}
\def\Z{{\bf Z}}
\def\z{{\bf z}}

\def\bzero{{\bf 0}}

\def\bq{\begin{equation}}
\def\eq{\end{equation}}

\def\wh{\widehat}

\def\trans{^{\rm T}}

\def\squarebox#1{\hbox to #1{\hfill\vbox to #1{\vfill}}}
\def\btheta{{\boldsymbol \theta}}
\def\1{{\bf 1}}

\def\var{\hbox{var}}

\def\bse{\begin{eqnarray*}}
\def\ese{\end{eqnarray*}}
\def\be{\begin{eqnarray}}
\def\ee{\end{eqnarray}}
\def\bsq{\begin{equation*}}
\def\esq{\end{equation*}}
\def\bq{\begin{equation}}
\def\eq{\end{equation}}
\def\wh{\widehat}

\def\trans{{\mathrm{T}}}
\def\boxit#1{\vbox{\hrule\hbox{\vrule\kern6pt\vbox{\kern6pt#1\kern6pt}\kern6pt\vrule}\hrule}}

\def\boxit#1{\vbox{\hrule\hbox{\vrule\kern6pt
          \vbox{\kern6pt#1\kern6pt}\kern6pt\vrule}\hrule}}

\catcode`@=11
\def\@evenhead{\vbox{\hbox to\textwidth{\tiny \hfill \hfill \today } }}
\def\@oddhead{\vbox{\hbox to \textwidth{\tiny \hfill \hfill \today } }}

\begin{document}

\def\spacingset#1{\renewcommand{\baselinestretch}%
{#1}\small\normalsize} \spacingset{1}
{
\renewcommand{\baselinestretch} {1}
\title{  \textbf{{High-Dimensional Covariate-Augmented Overdispersed Multi-Study Poisson  Factor Model
}}}

\if1\blind
\author{Wei Liu$^{1}$, Qingzhi Zhong$^2*$\\
{\small $^1$School of Mathematics, Sichuan University, Chengdu, China}\\
{\small$^2$School of Economics, Jinan University, Guangzhou, China}
}
\fi

\maketitle
\thispagestyle{empty}

\if1\blind
\begin{singlespace}
\begin{footnotetext}
{*Corresponding author.   Email: \emph{zhongqz19@icloud.com}.}
\end{footnotetext}
\end{singlespace}
\fi

\bigskip
\begin{abstract}
Factor analysis for high-dimensional data is a canonical problem in statistics and has a wide range of applications. However, there is currently no factor model tailored to effectively analyze high-dimensional count responses with corresponding covariates across multiple studies, such as the single-cell sequencing dataset from a case-control study. 
In this paper, we introduce factor models designed to jointly analyze multiple studies by extracting study-shared and specified factors. Our factor models account for heterogeneous noises and overdispersion among counts with augmented  covariates. We propose an efficient and speedy variational estimation procedure for estimating model parameters, along with a novel criterion for selecting the optimal number of  factors and the rank of regression coefficient matrix. The consistency and asymptotic normality of estimators are systematically investigated by connecting variational likelihood and profile M-estimation. Extensive simulations and an analysis of a single-cell sequencing dataset are conducted to demonstrate the effectiveness of the proposed multi-study Poisson factor model.
\end{abstract}

\noindent%
{\it Keywords:} Multi-study; Count data; Variational inference;  High-dimensional Factor analysis.
\vfill

\newpage
\spacingset{1.9} 
\section{Introduction}

In both scientific research and practical applications, the integration of data from diverse sources or studies holds paramount significance~\citep{korsunsky2019fast}. This process plays a crucial role in fortifying the robustness of findings, alleviating bias and error, and broadening the scope of external validity~\citep{argelaguet2020mofa+,liu2023probabilistic}. When dealing with multiple studies, a central consideration involves preserving shared commonalities across studies while discerning variations specific to each, given their inherent heterogeneity. To address this challenge, multi-study factor analysis model (MSFA)~\citep{de2019multi,de2021bayesian} have demonstrated a potent ability to integrate information from multiple studies. 
This methodology ensures that common features are preserved in the study-shared factors, while study-specified factors distinguish unique characteristics specific to each study, proving advantageous for various downstream tasks in real-world scenarios~\citep{argelaguet2020mofa+}. However, it's essential to note that the current MSFA assumes a linear relationship between variables, which may be easily violated in practical situations, such as with count-type data.

Count-type data are increasingly prevalent in biological, life and medical science attributed to advancements in sequencing technology. For example, the single-cell RNA sequencing (scRNA-seq) technologies~\citep{liu2022joint} and multi-modal sequecing technologies~\citep{lakkis2022multi} produce a massive of count data consisting of read counts of genes or other biological molecule measured for each cell from different tissues. Moreover, both the number of variables ($p$) and the sample size ($n$) within datasets grow substantially, even the number of variables exceeds the sample size due to the increase of throughout and resolution of sequencing technology.  
Moreover, the overdispersion phonemena~\citep{liang1993case} are often observed in count data of interest, and there may be extra variables of potential relevance that accompany those of primary interest.   For instance, consider the motivating PBMC data from a case-control study measured through CITE-seq technology. Each dataset of case and control groups comprises a cell-by-gene count matrix with thousands of cells and tens of thousands of genes. Researchers are interested in the common features and difference in the gene expression pattern between these two groups. Additionally, it includes hundreds of protein markers that account for gene expression levels in each group. In the framework of MSFA, to accommodate additional covariates,  \cite{de2023multi} proposed multi-study factor regression model (MSFR) by adding a regression term in MSFA. However, MSFR is not well-suited for handling count-type data and is inadequate for addressing scenarios characterized by a large number of variables and a small sample size.

Numerous methods in the literature address count data within the latent factor model framework, primarily designed for analyzing data from a single source or study. For instance, \cite{btt091} and \cite{xu2021zero} introduced the Poisson factor model, modeling the dependence of count data on latent factors through a log-linear model. The Generalized Factor Model~\citep{liu2023generalized} extends this framework to accommodate mixed-type variables using the exponential family, with count variables considered as a special case. To incorporate covariates, \cite{chiquet2018variational} and \cite{hui2017variational} proposed the probabilistic Poisson PCA model and the generalized linear latent variable model, respectively. However, these models face limitations in handling high-dimensional covariates and are unable to account for potential overdispersion not explained by latent factors. Recently, \cite{liu2024highdimensional} addressed these limitations by proposing covariate-augmented overdispersed Poisson factor models (COAP), introducing an additional error term in the log-linear model. When count data originate from multiple studies, existing methods lack the capability to capture study-specified features. 

To effectively analyze count data from multiple studies and address real-world requirements, an optimal factor model should possess the following capabilities: (a) handling data heterogeneity across sources; (b) effectively addressing the count nature of the data; (c) managing high-dimensional count variables; (d) accommodating overdispersion; and (e) incorporating additional covariates. To amalgamate these capacities, we introduce a multi-study covariate-augmented overdispersed Poisson factor model, referred to as MultiCOAP. Within MultiCOAP, we connect count variables from each study with study-shared factors, study-specified factors and additional covariates, along with a study-specified error term, through a log-linear model. This approach facilitates capturing the desired homogeneity of studies through shared factors while accounting for the intended heterogeneity among studies using study-unique factors. Additionally, it effectively captures heterogeneous noises and overdispersion not explained by factors through study-specified errors. Furthermore, MultiCOAP accommodates both low and high-dimensional covariates, enhancing its flexibility.
In addition to the mentioned contributions, we provide a theoretical investigation into the model identifiability, essential for uniquely determining parameters in models and ensuring interpretability. Thirdly, we propose a variational EM (VEM) algorithm that not only estimates model parameters but also provides posterior estimations of factors along with their posterior covariance, crucial for quantifying the uncertainty associated with estimated factors. Theoretically, we prove the convergence of the VEM algorithm and establish the consistency and asymptotic normality of the corresponding estimators within the variational likelihood framework. Fourth, we develop a criterion based on cumulative explained variance proportion to determine the number of factors under the MultiCOAP framework. Finally, numerical experiments demonstrate that MultiCOAP outperforms existing methods in both estimation accuracy and computational efficiency.

The subsequent sections of the paper are structured as follows. In Section \ref{sec:model}, we offer an overview of the model setup and estimation method for MultiCOAP. Section \ref{sec:asymp} is dedicated to establishing the convergence rate and asymptotic normality of estimators. Following that, we detail the variational EM algorithm for MultiCOAP in Section \ref{sec:alg}, along with presenting the model selection method and extensions in Section \ref{sec:modelselect}. To assess the performance of MultiCOAP, simulation studies are conducted in Section \ref{sec:simu}, and real data analysis is presented in Section \ref{sec:real}. In Section \ref{sec:dis}, we briefly discuss potential avenues for further research in this field. Technical proofs and additional numerical results are provided in the Supplementary Materials. Furthermore, we have seamlessly integrated MultiCOAP into an efficient and user-friendly R package, which is conveniently accessible at \url{https://github.com/feiyoung/MultiCOAP}.

\nvs

\section{Model and estimation} \label{sec:model}
We assume there are $S$ data sources from different studies. In each source $s\in \{1,\cdots,S\}$, we observe a count matrix $\X_s=(x_{sij})_{n_s \times p}$ and additional covariate matrix $\Z_s=(z_{sij})_{n_s \times d}$. To model this type of data,
we propose a multi-study covariate-augmented overdispersed Poisson factor  model:
\nvs
\begin{eqnarray}
  &&x_{sij}|\tilde y_{sij} \sim  Pois(a_{si} \tilde y_{sij}), \label{eq:xij}\\
  && y_{sij} = \ln \tilde y_{sij} = \z_{si}^\trans \bb_j +  \f_{si}^\trans\ba_j + \h_{si}^{\trans}\bg_{sj}+ \varepsilon_{sij}, \label{eq:yij}
\end{eqnarray}
where, for the $i$th individual from the study $s$, $a_{si}$ is the known normalization factor, $\widetilde   y_{sij}$  and $y_{sij}$ are the unobservable Poisson rate and its logarithm  representing the underlying value and log-value of the variable $j$, respectively. The Poisson rate depends on the covariates $\z_{si}=(z_{si1}, \cdots, z_{sid})^{\trans}$, the study-shared latent factor $\f_{si}=(f_{si1}, \cdots, f_{siq})^{\trans}$ and study-specified latent factor $\h_{si}=(h_{si1}, \cdots, h_{siq_s})^{\trans}$ via a log-linear model. $\bb_j$ is a $d$-dimensional vector
of coefficients, including an intercept that represents the mean log-expression of variable $j$,
together with $d -1$ coefficients corresponding to the associated covariates. $\f_{si}$ and $\h_{si}$ accounts for both unknown or unavailable covariates from $\z_{si}$,  and nonlinear correlation among variables $\x_{si}$. $\ba_j$ and $\bg_{sj}$ are the loading vectors for variable $j$, corresponding to the study-shared factor and study-specified factor, respectively. Let $\I_q$ be the $q$-by-$q$ identity matrix.  We assume $\varepsilon_{sij}\sim N(0, \lambda_{s})$, $\f_{si} \sim N(\bzero, \I_q)$, $\h_{si} \sim N(\bzero, \I_{q_s})$, and  $\f_{si}, \h_{si}$ and  $\bvarepsilon_{si}$ are mutually independent with $\bvarepsilon_{si}=(\varepsilon_{si1},\cdots,\varepsilon_{sip})^{\trans}$. Different from COAP \citep{liu2024highdimensional},  we assume distributions for the latent factor while COAP regards the latent factor as a infinite parameter and directly estimate it. There are two advantages by exerting the distributions. First, it reduces the model parameters significantly. Second, by the distribution assumption, the uncertainty of posterior estimation for $\f_{si}$ and $\h_{si}$ can be measured.
\nvs
\subsection{Identifiability conditions}
The issue of identifiability is a common challenge in latent factor models. Therefore, we begin by investigating identifiable conditions to ensure the meaningful interpretation of models \eqref{eq:xij} and \eqref{eq:yij}.
Let $q_{\min}=\min_s q_s$, $\A=(\ba_1, \cdots, \ba_p)^{\trans} \in \mathbb{R}^{p\times q}$ and $\B_s=(\bg_{s1}, \cdots, \bg_{sp})^{\trans} \in \mathbb{R}^{p \times q_s}$ and $\bb=(\bb_1, \cdots, \bb_p)^{\trans} \in \mathbb{R}^{p \times d}$. The uniqueness of determining the parameters is achieved by imposing a set of identifiability conditions:
\begin{itemize}
  \item[(A1)] $\z_{si} \perp (\f_{si}, \h_{si},  \bvarepsilon_{si})$ and $\mathrm{E}(\z_{si}\z_{si}^{\trans})$ is positive definite.

  \item[(A2)] $(\A,\B_1)^{\trans}(\A, \B_1)$ and $\{\B_s^{\trans}\B_s, 2\leq s\leq S\}$ are all diagonal matrices with the decreasing diagonal positive entries, and for $\A$ and $\B_s, s\leq S$, the first nonzero  element of each column  is positive.
  \item[(A3)] $p-1> q+q_s$; 
  \item[(A4)] For $1\leq k\leq q_{\min}$,  there exist $s_1$ and $s_2$ such that $\B_{s_1,. k}\neq \B_{s_2, .k}$, where $\B_{s_1, .k}$ is the $k$-th column of $\B_{s_1}$.
\end{itemize}
Condition (A1) guarantees the identifiability of $\bb$ and signifies that the factors $(\f_{si}, \h_{si})$ extract information from $\x_{si}$ not explained by $\z_{si}$.
Condition (A2) prevents the rotation invariance of $\A$ and $\B_s$.
Condition (A3) ensures that the number of study-shared and study-specified latent factors is less than the number of count variables minus one. This condition is typically satisfied.
Condition (A4) ensures that $h_{sik}$ is indeed the study-specified factor. Otherwise, if $\B_{s, .k}$ is the same across all studies ($s \leq S$), then it also becomes a study-shared factor.

We state the identifiability in the following theorem, with its proof deferred to the  Supplementary Materials.
\begin{Theorem}
 Under Conditions (A1)--(A4), models \eqref{eq:xij} and \eqref{eq:yij} are identifiable.
\end{Theorem}

\nvs
\subsection{Estimation}
Let $\btheta=(\btheta_s, s\leq S)$ be the collection of all model parameters with $\btheta_s=(\bb, \A, \B_s, \lambda_s)$, $\u_{si}=(\x_{si}^{\trans}, \z_{si}^{\trans})^{\trans}$ with $\x_{si}=(x_{si1},\cdots,x_{sip})^{\trans}$,  and $l(\btheta_s;\u_{si})$ be the log-likelihood function for the $i$-th observation in the study $s$. After dropping a constant, we  write $l(\btheta_s;\u_{si})$ as
\nvs
$$l(\btheta_s;\u_{si}) = \ln \int P(\x_{si}|\y_{si}) P(\y_{si}|\f_{si}, \h_{si}, \z_{si}) P(\f_{si})P(\h_{si})d\y_{si}d\f_{si} d\h_{si}.$$
Let $l(\btheta)=\sum_{s=1}^{S}\sum_{i=1}^{n_s} l(\btheta_s;\u_{si})$. The maximum likelihood estimate of $\btheta$
is the maximizer of $l(\btheta)$. However, estimation and inference
associated with $l(\btheta)$ are challenging due to the fact that the integral in
$l(\btheta)$ is analytically intractable.

Motivated by \cite{westling2019beyond}, we  approximate the maximum likelihood by utilizing a variational approximation. Let $g_{si}$ be an arbitrary
density function of the latent random vector $(\y_{si}, \f_{si}, \h_{si})$ and define
\vspace{-0.3in}
$$ \tilde l(\btheta_s, g_{si};\u_{si})= \int \ln \left(\frac{P(\x_{si}|\y_{si}) P(\y_{si}|\f_{si}, \h_{si}, \z_{si}) P(\f_{si})P(\h_{si})}{g_{si}(\y_{si}, \f_{si}, \h_{si})}\right) g_{si}(\y_{si}, \f_{si}, \h_{si})d\y_{si}d\f_{si} d\h_{si} $$
By Jensen’s inequality, we have
\nvs
$$l(\btheta_s; \u_{si}) \geq \tilde l(\btheta_s, g_{si}; \u_{si})$$
with equality if and only if $g_{si}$ is the posterior density of $(\y_{si}, \f_{si}, \h_{si})$ given
$\x_{si}$ and $\z_{si}$.
$\tilde l(\btheta_s, g_{si};\u_{si})$ is the corresponding variational lower bound on $l(\btheta_s;\u_{si})$.
Let $g = \Pi_{s,i} g_{si}$ and
\nvs
$$\tilde l(\btheta, g)=\sum_{s,i} \tilde l(\btheta_s, g_{si};\u_{si}). $$
We have $l(\btheta)\geq \tilde l(\btheta, g)$. Variational approximation aims to
maximize $\tilde l(\btheta, g)$ over a distribution family, i.e., $g\in \mathcal{G}$, so as to find the best variational density
that is closest in Kullback–Leibler divergence to the posterior
density. This leads to the optimization problem $\hat g = \arg\max_{g\in \mathcal{G}} \tilde l(\btheta, g)$. The variational lower bound $l(\btheta, \hat g)$ is known as the variational
log-likelihood~\citep{wang2019frequentist}. We then maximize $l(\btheta, \hat g)$ with respect to $\btheta$, i.e.,
$\wh\btheta= \arg\max_{\btheta} \tilde l(\btheta, \hat g)$. $\wh\btheta$ is called the maximum variational likelihood estimate.

To make $\hat g$ tractable, we assume a variational mean field family, i.e.,
$g_{si}(\y_{si}, \f_{si}, \h_{si})=\Pi_{j} N(y_{sij}; \mu_{sij}, \sigma^2_{sij}) \times N(\f_{si}; \m_{f,si}, \S_{f,si})\times N(\h_{si}; \m_{h,si},\S_{h,si})$, where  $N(y; \mu, \sigma^2)$ is the (multivariate) normal density function of $y$ with mean (vector) $\mu$ and variance/covariance $\sigma^2$.
Let $\bxi_{si} = \{\mu_{sij}, \sigma^2_{sij}, \m_{f,si}, \S_{f,si}, $ $\m_{h,si},\S_{h,si}, j\leq p\}$ and $\bxi=\{\bxi_{si}, i\leq n_s,  s \leq S\}$ that is the collection of variational parameters.

Then, $\tilde l(\btheta_s, g_{si}; \u_{si})$ becomes
\nvs
\begin{eqnarray}\label{eq:elbosi}
  \tilde l_{si}(\btheta_s, \bxi_{si}) &=& \sum_{j} \bigg\{x_{sij}\mu_{sij} -  a_{si}\exp(\mu_{sij}+\sigma_{sij}^2/2) - \frac{1}{2}\big\{(\mu_{sij}-\z_{si}^\trans \bb_j -\ba_j^\trans\m_{f,si} - \bg_{sj}^{\trans} \m_{h,si})^2/\lambda_{s}\nonumber\\
  &&  + \lambda_{s}^{-1}(\sigma_{sij}^2 + \ba_j^\trans \S_{f,si}\ba_j + \bg_{sj}^{\trans} \S_{h,si} \bg_{sj}) + \ln \lambda_{s} \big\} \bigg\} - \frac{1}{2}\big\{ \m_{f,si} ^\trans \m_{f,si}  + \mathrm{Tr}(\S_{f,si}) \nonumber\\
    && + \m_{h,si} ^\trans \m_{h,si}  + \mathrm{Tr}(\S_{h,si})  \big\} + \frac{1}{2}\big(\sum_j \ln \sigma_{sij}^2 + \ln |\S_{f,si}| + \ln |\S_{h,si}|\big) + c,
\end{eqnarray}
where $\mathrm{Tr}$ and $|\bullet|$  represent the trace and determinant operators, respectively, for a square matrix and $c$ is a constant independent of $(\btheta_s, \bxi_{si})$.

By maximizing the variational lower bound $\tilde l(\btheta, \bxi) = \sum_{s,i} \tilde l_{si}(\btheta_s, \bxi_{si})$, we derive the estimators $\wh\btheta$ and $\wh\bxi$. Among these, the posterior means $\wh\m_{f,si}$ and $\wh\m_{h,si}$ within $\wh\bxi$ serve as the posterior estimates for $\f_{si}$ and $\h_{si}$—the low-dimensional representations of $\x_{si}$ in study-shared and study-specified aspects, respectively. Simultaneously, the estimation uncertainty associated with $\f_{si}$ and $\h_{si}$ is encapsulated by $\wh\S_{f,si}$ and $\wh\S_{h,si}$.
Subsequently, we delve into the theoretical properties of the estimators for model parameters and expound upon the practical implementation algorithm.
\nvs
\section{Asymptotical analysis}\label{sec:asymp}
Recent studies have explored the asymptotic properties of the variational likelihood estimator \citep{peterAsy, westling2019beyond, pang2023factor}. Our research is distinct from these results in two key aspects. Firstly, our results are presented at a multi-study level rather than focusing on a single study, necessitating distinct  technical handling. Secondly, our emphasis lies in exploring the asymptotic properties of both study-shared and study-specified parameters. To establish consistency and asymptotic normality, we employ M-estimation on the profiled variational log-likelihood. The following Theorem elucidates the relationship between maximum variational estimation and M-estimation.

\begin{Theorem}
Let $\rho_p(\btheta_s; \x_{si}, \z_{si})=\tilde l_{si}(\btheta_s, \bxi_{si}(\btheta_s))$. Then, (i) for all $\btheta_s$, $\bxi_{si}\mapsto \tilde l_{si}(\btheta_s, \bxi_{si})$ possesses a unique maximizer $\bxi_{si}(\btheta_s)=\arg\max_{\bxi_{si}} \tilde l_{si}(\btheta_s, \bxi_{si})$, (ii) the maximum variational likelihood estimator $\wh\btheta$ is equal to the  maximizer of the profiled variational log-likelihood $\sum_{s,i}\rho_p(\btheta_s; \x_{si}, \z_{si})$.
\end{Theorem}

 Let $D_\bullet,D^2_\bullet,\lambda_{\min}(\bullet),\|\bullet\|_{\max}$ denote the first, second derivative, smallest eigenvalue and max norm operators with respect to $\bullet$, and $\|\bullet\|$ denote either the Euclidean norm for a vector or the Frobenius matrix norm for a matrix. Denote $\r_n=(\sqrt{\sum_{s}n_s}\1^{\trans}_{pq+pd},\sqrt{n_s}\1^{\trans}_{pq_s+1},s\leq S)^{\trans},\K_1^{-1}= \mathrm{E}\{-D^2_{\btheta\btheta}\sum_{s,i}\rho_p(\btheta_{s0}; \x_{si}, \z_{si})\}$, and $1/{\r_n}$ is the vector obtained by element-wise division of $\r_n$ by 1.  For a column vector $\a$, we define $\mathrm{diagmat}(\a)$ as an operator that transforms $\a$ into a diagonal matrix. Then, we impose the following assumptions.
\begin{description}
\item [(C1)] (A1)--(A4) hold.
\item [(C2)] There exists some positive constant $M$ such that $\|\btheta\|_{\max}\leq M$ and $\mathrm{E}\|\z_{si}\|^2\leq M$.
\item [(C3)]  There exists positive constant $M$ such that $\lambda_{\min}\{\mathrm{diagmat}(1/\r_n)\K_1^{-1}\mathrm{diagmat}(1/\r_n)\}>M$.
\end{description}
Condition (C1) ensures the model identifiability. Condition (C2) assumes the model parameters and the second-order moment of $\z_{si}$ are bounded. Condition (C3) guarantees that the profiled variational objective function is concave in the population level. In the context of a single study, Condition (A3) in \cite{pang2023factor} shares similarities with Condition (C3). However, it is assumed to be applicable at the sample level, making it challenging to meet in practical scenarios.


Then, we present the convergence rates of the estimators of the study-shared  and study-specified parameters.
\begin{Theorem}\label{thm3}
Suppose $\btheta_0=\arg\max_{\btheta}\mathrm{E}\sum_{s,i}\rho_p(\btheta_s; \x_{si}, \z_{si})$. If $p^4=o(\min_sn_s)$ and Conditions (C1)-(C3) hold, then
\nvs
\begin{eqnarray*}
&&\|\widehat\bb-\bb_0\|^2+\|\widehat\A-\A_0\|^2=O_p(\frac{p}{\sum_sn_s}),\\
&& \|\widehat\B_s-\B_{s0}\|^2+|\widehat\lambda_s-\lambda_{s0}|^2=O_p(\frac{p}{n_s}).
 \end{eqnarray*}
\end{Theorem}
Theorem \ref{thm3} reveals distinct convergence rates for the estimated study-shared parameters and study-specified parameters, characterized by  $O_p(\frac{p}{\sum_sn_s})$ and $O_p(\frac{p}{n_s})$, respectively. This aligns with the fact that study-shared parameters are estimated using observations from all studies, while study-specified parameters are estimated solely from observations within a single study. Next, we present the asymptotical normality of estimators.

\begin{Theorem}\label{thm4}
Under the conditions in Theorem \ref{thm3}, it holds that
\nvs
$$\widehat\btheta-\btheta_0=\K_1^{-1}D_{\btheta}\sum_{s,i}\rho_p(\btheta_{s0}; \x_{si}, \z_{si})+o_p(1/\r_n).$$
Let $\G=(\mathrm{diagmat}(\r_n)\K_1^{-1}\K_2\K_1^{-1}\mathrm{diagmat}(\r_n))$, where $\K_2=\mathrm{E}\{D_{\btheta}\sum_{s,i}\rho_p(\btheta_{s0}; \x_{si}, \z_{si})$ \\ $D^{\trans}_{\btheta}\sum_{s,i}\rho_p(\btheta_{s0}; \x_{si}, \z_{si})\}$. Denote $\G_{\bb_j\bb_j},\G_{\ba_j\ba_j},\G_{\bg_{sj}\bg_{sj}}$ and $\G_{\lambda_s\lambda_s}$ are submatrices of $\G$ corresponding to $\bb_j,\ba_j,\bg_{sj}$ and $\lambda_s$, respectively.  Then, it holds that
\nvs
 \begin{eqnarray*}
&& \sqrt{\sum_s n_s}(\widehat\bb_j-\bb_{j,0})\rightarrow N(\0,\G_{\bb_j\bb_j}),\\
&&\sqrt{\sum_s n_s}(\widehat\ba_j-\ba_{j,0})\rightarrow N(\0,\G_{\ba_j\ba_j}),\\
&&\sqrt{n_s}(\widehat\bg_{sj}-\bg_{sj,0})\rightarrow N(\0,\G_{\bg_{sj}\bg_{sj}}),\\
&&\sqrt{n_s}(\widehat\lambda_s-\lambda_{s,0})\rightarrow N(\0,\G_{\lambda_s\lambda_s}).
  \end{eqnarray*}
\end{Theorem}
In contrast to \cite{pang2023factor}, we alleviate the dimensionality constraint on count variables for establishing the asymptotic normality of the variational likelihood estimator. Specifically, \cite{pang2023factor} considered a model for a single study with a sample size of $n$ and imposed a stringent condition of $p^5=o(n)$ due to their independent bounding of the infinite-dimensional norm term and the remainder of the second-order Taylor expansion of the score function \citep{pang2023factor}. In contrast, we provide a refined upper bound by bounding the product of these two terms. Consequently, our framework adopts a more relaxed condition, requiring  only $p^4=o(\min_s n_s)$  as detailed in Theorem \ref{thm4}; see Appendix D in Supplementary Materials for further details.
\section{Algorithm}\label{sec:alg}
Let $\X=(\x_{11},\x_{12},\cdots,\x_{Sn_S})^\trans$ and define $\Y,\F,\H$ and $\Z$ similarly. Denote $\F_s=(\f_{s1},\cdots,\f_{sn_s})^\trans \in \mathbb{R}^{n_s \times q}$ and $\H_s=(\h_{s1},\cdots,\h_{sn_s})^\trans \in \mathbb{R}^{n_s \times q_s}$.
By equation \eqref{eq:elbosi}, the variational evidence lower bound (ELBO) of $l(\btheta)$ is given by
\nvs
\begin{eqnarray}
  \tilde l(\btheta,\bxi) &=& \mathrm{E}_g  \ln P(\X, \Y, \F, \H|\Z) - \mathrm{E}_g \ln g(\Y,\F, \H) \nonumber\\
  &=& \sum_{s,i,j} \left\{x_{sij}\mu_{sij} -  a_{si}\exp(\mu_{sij}+\sigma_{sij}^2/2) - \frac{1}{2}\{(\mu_{sij}-\z_{si}^\trans \bb_j -\ba_j^\trans\m_{f,si} - \bg_{sj}^{\trans} \m_{h,si})^2/\lambda_{s}\right. \nonumber\\
  && \left.  + \lambda_{s}^{-1}(\sigma_{sij}^2 + \ba_j^\trans \S_{f,si}\ba_j + \bg_{si}^{\trans} \S_{h,si} \bg_{sj}) + \ln \lambda_{s} \} \right\} - \sum_{s,i} \frac{1}{2}\{ \m_{f,si} ^\trans \m_{f,si}  + \mathrm{Tr}(\S_{f,si}) \nonumber\\
    && + \m_{h,si} ^\trans \m_{h,si}  + \mathrm{Tr}(\S_{h,si})  \} + \frac{1}{2}\sum_{s,i}(\sum_j \ln \sigma_{sij}^2 + \ln |\S_{f,si}| + \ln |\S_{h,si}|) + c , \label{eq:elbo}
\end{eqnarray}
where $\mathrm{E}_g$ denotes taking expectation with respect to $(\Y, \F, \H)$ using the assumed variational distribution $g$. Based on this ELBO function, we formulate a variational EM algorithm.

Denote $f_{sij}(y)=x_{sij}y- a_{si}\exp(y) - \frac{1}{2} \{(y-\z_i^\trans \bb_j-\ba_j^{\trans} \m_{f,si}-\bg_j^\trans\m_{h,si})^2 + \ba_j^\trans \S_{f,si} \ba_j + \bg_j^\trans \S_{h,si} \bg_j \}/\lambda_{s}$. In the E-step, we transform the intractable posterior expectation of latent variables into a tractable optimization problem concerning the variational parameters. Drawing inspiration from the methodology detailed in \cite{liu2024highdimensional}, we employ the Laplace approximation based on $f_{sij}(y)$ and utilize a Taylor approximation for $\exp(y)$ within $f_{sij}(y)$. This yields the iterative expressions for $\mu_{sij}$ and $\sigma_{sij}^2$ as follows:
\nvs
\begin{equation}\label{eq:updateY}
 \mu_{sij}=\frac{x_{sij} - a_{si} \exp(y_0)(1-y_0)+\lambda_{s}^{-1}\tilde z_{sij}}{\lambda_{s}^{-1} +a_{si} \exp(y_0)},  \,\  \sigma_{sij}^2=\frac{1}{a_{si}\exp(\mu_{sij}) + \lambda_{s}^{-1}}.
\end{equation}
where $y_0$ is the previous iterative value of $\mu_{sij}$ and $\tilde z_{sij}=\z_{si}^\trans \bb_j + \ba_j^\trans\m_{f,si} + \bg_{sj}^{\trans} \m_{h,si}$.

Next, we update the variational parameters of $\f_{si}$ and $\h_{si}$. Differentiating $\tilde l(\btheta,\bxi)$ with respect to (w.r.t.) $(\m_{f,si}, \S_{f,si})$ and $(\m_{h,si}, \S_{h,si})$, and set the derivatives to zeros, we obtain
\nvs
\begin{eqnarray}
   && \S_{f,si} = (\A^\trans \A / \lambda_s+ \I_q)^{-1}, \label{eq:Sfi}\\
   && \m_{f,si} = \S_{f,si} \sum_j (\mu_{sij} -\z_{si}^\trans \bb_j-\bg_{sj}^{\trans}\m_{h,si})\lambda_{s}^{-1} \ba_j, \label{eq:mufhi} \\
    && \S_{h,si} = (\B_s^\trans  \B_s/ \lambda_s + \I_{q_s})^{-1}, \label{eq:Shi}\\
   && \m_{h,si} = \S_{h,si} \sum_j (\mu_{sij} -\z_{si}^\trans \bb_j-\ba_{j}^{\trans}\m_{f,si})\lambda_{s}^{-1} \bg_{sj}. \label{eq:muhi}
\end{eqnarray}
In the M-step, differentiating $\tilde l(\btheta,\bxi)$ w.r.t. model parameters, and setting the derivatives to zeros, we have
\nvs
\begin{eqnarray}
   \ba_j &=& \{ \sum_{s,i}(\m_{f,si}\m_{f,si}^{\trans} + \S_{f,si})\}^{-1} \sum_{s,i} \m_{f,si}(\mu_{sij} - \z_{si}^\trans \bb_j - \bg_{sj}^{\trans} \m_{h,si}), \label{eq:updata_bj} \\
   \bg_{sj} &=& \{ \sum_{i}(\m_{h,si}\m_{h,si}^{\trans} + \S_{h,si})\}^{-1} \sum_{i} \m_{h,si}(\mu_{sij} - \z_{si}^\trans \bb_j - \ba_{j}^{\trans} \m_{f,si}), \label{eq:updataSigma} \\
   {\lambda_s}&=&  \frac{1}{n_s p} \sum_{i,j} \left\{(\bar y_{sij} -\z_{si}^{\trans}\bb_j)^2+  \sigma_{sij}^2 + \ba_j^{\trans} \S_{f,si}\ba_j + \bg_{sj}^{\trans} \S_{h,si} \bg_{sj}\right\}, \\
   \tilde\bb&=&\bar\Y^{\trans}\bar\Z(\bar\Z^{\trans}\bar\Z)^{-1}, \label{eq:updatabb}
\end{eqnarray}
where $\bar y_{sij}=\mu_{sij}-  \ba_{j}^{\trans} \m_{f,si}- \bg_{sj}^{\trans} \m_{h,si}$, $\bar\Y = (\bar\Y_1^{\trans}/\sqrt{\lambda_s}, \cdots, \bar\Y_S^{\trans}/\sqrt{\lambda_s})^{\trans}$ with $\bar \Y_s$ being the $n_s$-by-$p$ matrix with $(i,j)$-th entry $\bar y_{sij}$,  and $\bar\Z=(\Z_1^{\trans}/\sqrt{\lambda_s}, \cdots, \Z_S^{\trans}/\sqrt{\lambda_s})^{\trans}$.
We summarize the implementation of MultiCOAP in Algorithm 1, with its convergence rigorously established in Proposition 1, both detailed in the Appendix B of Supplementary Materials.

\section{Model selection}\label{sec:modelselect}
\nvs
\subsection{Determining the number of factors}\label{sec:select}
To determine the numbers of study-shared factors ($q$) and study-specified factors ($q_s$), we employ a method based on \underline{cu}mulative explained variance \underline{p}roportion, simplified as 'CUP' criterion. Define $\nu_{f,k}= \sum_{j=1}^{p} \alpha_{jk}^2$ and $\nu_{h,sk}=\sum_{j=1}^{p} \gamma_{sjk}^2$, where $\alpha_{jk}$ and $\gamma_{sjk}$ are the $k$-th element of $\ba_j$ and $\bg_{sj}$, respectively. Considering $\var(\f_{si})=\I_q$, we have $\var(\alpha_{jk}f_{sik})=\alpha_{jk}^2$, indicating the explained variance of the $j$-th variable by the $k$-th study-shared factor. Thus, $\nu_{f,k}=\sum_{j=1}^{p} \alpha_{jk}^2$ is the total explained variance by the $k$-th factor. Given an upper bound of $q$, i.e., $q_{max}$, we seek $\hat q$ such that $\hat q = \min\{q: \frac{\sum_{k=1}^{q}\nu_{f,k}}{\sum_{k=1}^{q_{max}}\nu_{f,k}} > \tau\}$, where $\tau$ is the desired variance proportion, i.e., $\tau=95\%$. The underlying logic is that if factors with indices greater than  $\hat q$ are not attributed to the study-shared variations,  then the cumulative variance proportion of the first $\hat q$ factors is sufficiently high. Applying the same principle, we determine $\hat q_s$ as $\hat q_s = \min\{q_s: \frac{\sum_{k=1}^{q_s}\nu_{h,sk}}{\sum_{k=1}^{q_{s,max}}\nu_{h,sk}} > \tau\}$, where $q_{s,max}$ is the upper bound of $q_s$.
In practical implementation, we substitute $\nu_{f,k}$ and $\nu_{h,sk}$ with their respective estimators.

\nvs

\subsection{Regularization for the regression coefficient matrix}
Here, we expand our estimation of $\bb$ to incorporate regularization. A reasonable choice for regularization involves imposing a low-rank structure~\citep{she2017robust,liu2024highdimensional}, given the inherent  high correlation among $\x_{si}$ and possible correlation among $\z_{si}$. Incorporating a low-rank constraint provides at least two advantages. Firstly, it allows us to leverage the correlation information among observed variables  in estimating $\bb$. Secondly, in situations where the number of covariates ($d$) grows disproportionately to the sample size ($n$), the consistency of the  estimator   may be compromised without structural constraints.
To obtain the update of $\bb$, we  solve the following constrained optimization:
\nvs
{\begin{equation}\label{eq:matBeta2}
\min_{\bb} \mathrm{Tr}\left\{\frac{1}{n}(\bar{\Y}-\bar\Z\bb^{\trans})^{\otimes 2}\right\}, \mbox{ subject to rank}(\bb)\leq r,
\end{equation}
which reduces to the reduced-rank regression problem~\citep{izenman1975reduced}, where $\K^{\otimes 2}=\K^{\trans} $} for any matrix $\K$. The closed form is given by
\nvs
\begin{eqnarray} \label{eq:betasep}
     && \bb = \bar \V_{(r)} \bar \V_{(r)}^{\trans} \tilde\bb,
\end{eqnarray}
where $\bar \V_{(r)}$ is the matrix consisting of eigenvectors of $\tilde\bb(n^{-1}\bar\Z^{\trans}\bar\Z)\tilde\bb^{\trans}$ that correspond to its first $r$ largest eigenvalues, and $\tilde\bb$ is given in equation \eqref{eq:updatabb}. We simply replace \eqref{eq:updatabb} with \eqref{eq:betasep} when considering this regularization, while leaving the update of other parameters unchanged. To determine the underlying rank in practice, given an upper bound of $r_{max}$, we initially obtain the estimator $\wh\bb$ based on the rank $r_{max}$. Subsequently, we determine $\hat r$ as $\hat r= \min\{r: \frac{\sum_{k=1}^{r}\hat\nu_{\bb,k}}{\sum_{k=1}^{r_{max}}\hat\nu_{\bb,k}} > \tau\}$, where $\hat\nu_{\bb,k}$ the $k$-th largest eigenvalue of $\wh\bb^{\trans}\wh\bb$. The intuition behind this process is to identify the valid rank of the low-rank matrix $\wh\bb$ as the rank $\hat r$  sufficiently explains the information in $\wh\bb$.

\nvs
\section{Simulation}\label{sec:simu}
We conduct a comparative analysis between the proposed method and newly developed MSFR~\citep{de2023multi}, the implementation of which can be found at \url{https://github.com/rdevito/MSFR}. MSFR employs a linear model to capture the relationship between $\x_{si}$ and $(\z_{si}, \f_{si}, \h_{si})$. However, it overlooks the count nature of $x_{sij}$ when applied to count data. In addition to employing $x_{sij}$ as input for MSFR, we investigate the transformation of data using $\ln(1+x_{sij})$. This transformation serves to convert the data into a continuous format, aligning it more closely with the assumptions underlying MSFR. 
Simultaneously, we compare our method with COAP \citep{liu2024highdimensional}, which disregards the study-specified factors, and only estimates the loadings and factors shared across studies.

For two matrices $\wh\D$ and $\D$, we use trace statistic~\citep{doz2012quasi}, defined as $\mathrm{Tr}(\wh\D,\D)=\frac{\mathrm{Tr}\{\D^{\trans} \wh\D(\wh\D^{\trans}\wh\D)^{-1}\wh\D^{\trans}\D\}}{\mathrm{Tr}(\D^{\trans}\D)}$, to measure their similarity.  The trace statistic  ranges from 0 to 1, and it is considered better when it has a larger value.
We evaluate the accuracy of study-shared loading matrix $\A$ using $\mathrm{Tr}(\wh\A,\A_0)$, study-specified loading matrices, $\{\B_s\}_{s=1}^S$, using mean trace statistics $\frac{1}{S}\sum_{s=1}^{S}\mathrm{Tr}(\wh\B_s,\B_{s0})$, and assess the estimation accuracy of $\bb_0$ by employing mean estimation error (Er) $\sqrt{\frac{\|\wh\bb-\bb_0\|_F^2}{pd}}$.  In addition, we also care about the estimation of factor matrices, thus we evaluate the estimation performance of factor matrices, $\{\F_s\}_{s=1}^S$ and $\{\H_s\}_{s=1}^S$ using the mean trace statistics. We repeat 100 runs in all scenarios. 
\nvs
\subsection{Data generation}
We simulate data from models \eqref{eq:xij} and \eqref{eq:yij}. We set $a_{si}=1,\forall s,i$ by default, since we focus on the absolute level of $x_{sij}$. $\breve{\z}_{si}$ is independently generated from $N(\0_{d-1},(0.5^{|i-j|})_{(d-1)\times (d-1)}),$ \\$\z_{si} = (1, \breve{\z}_{si}^\trans)^\trans$, $\bb_0=4\rho_{z}\U_0 \V_0^{\trans}/p $ with $\U_0\in \mathbb{R}^{d\times r_0}$ and $\V_0 \in \mathbb{R}^{p\times r_0}$, where each element of $\U_0$ and $\V_0$ is drawn from standard Gaussian distribution, and $\rho_z$ is a scalar to control the signal strength. Next, we generate $\f_{si} \stackrel{i.i.d.}\sim  N(0, \I_q)$ and $\h_{si} \stackrel{i.i.d.}\sim  N(0, \I_{q_s})$. To generate $\A_0$ and $\B_{10}$, we first generate $\breve{\B}_1=(\breve{b}_{1jk})\in \mathbb{R}^{p\times (q+q_1)}$ with $\breve{b}_{1jk}\stackrel{i.i.d.}\sim N(0,1)$, then perform SVD, i.e., $\breve{\B}_1=\U_1 \bLambda_1 \V_1^{\trans}$ with the first nonzero entry of each column of $\U_1$ positive, and let $\bar\B_1= \rho_A \U_1 \bLambda_1$. Let $\A_0$ be the first $q$ columns of $\bar\B_1$ and $\B_{10}$ be the last $q_1$ columns of $\bar\B_1$. For $1<s \leq S$, we generate matrix $\breve{\B}_s\in \mathbb{R}^{p\times q_s}$ with SVD $\breve{\B}_s=\U_s \bLambda_s \V_s^{\trans}$ by the same way, and let $\B_{s0}=\rho_B \U_s \bLambda_s$. $\rho_A$ and $\rho_B$ are designed to control the signal strength of study-shared and study-specified factors, respectively. Note that $\A_0$ and $\{\B_{s0}\}_{s=1}^S$ satisfy the identifiable conditions (A2)--(A4) given in Section \ref{sec:model}. Finally, we set $\varepsilon_{sij} \stackrel{i.i.d.}\sim N(0, \sigma_0^2)$.
After generating $\bb_0, \sigma_0^2, \A_0$ and $\B_{s0}$, they are fixed in  repetition.  We set  $q=3, q_s=2,S = 2$, $\sigma_0^2=1$  and $d=10$ and $r=2$ without further specification. We evaluate the performance of MultiCOAP and other comparative methods through six examples. Additionally, we present another example, termed Example 7, involving high-dimensional covariates $\z$, which is discussed in Appendix E.1 of the Supplementary Materials.

\underline{\bf Example 1}. In this example, we examine the impact of sample size and the dimension of the count variable on the estimation of parameters of interest. We set $(\rho_A, \rho_B, \rho_z) = (2, 3.5, 0.1)$ and explore two cases. In Case 1, we hold $p$ constant at 100 while varying $(n_1, n_2)$ within $\{(50, 80), (100,200), (200, 300)\}$; in Case 2, we maintain $(n_1, n_2) = (100, 150)$ while altering $p$ within $\{50, 100, 150\}$. The results presented in Table \ref{tab:npvar1} demonstrate that MultiCOAP outperforms MSFR, MSFR-log, and COAP in both cases. Notably, MSFR, which neglects the count nature of variables, exhibits inferior performance. While log-transformation significantly improves parameter estimation of MSFR, this preprocessing may not accurately capture the underlying data generation mechanism, resulting in a substantial loss of statistical power compared to MultiCOAP. As depicted in Table \ref{tab:npvar1}, COAP exhibits lower performance compared to MultiCOAP and struggles notably in estimating study-specified factors and loadings. Furthermore, focusing on MultiCOAP, there is a conspicuous improvement in estimation accuracy for both study-shared and study-specified factors as the variable dimension expands from $50$ to $100$. However, as the parameter $p$ steadily increases to $150$, a slight diminishing trend in the accuracy of study-specified factor and loading estimations emerges, may attributed to the heightened complexity of estimation brought about by the increased dimension.
In contrast, a larger sample size enhances the estimation performance of loadings, factors and regression coefficient matrix, as it provides more information for estimating model parameters, thereby boosting factor estimation.

\begin{table}[h!]
    \centering\renewcommand\tabcolsep{3pt}
  \caption{Comparison of MultiCOAP and other competing models for parameter estimation. The reported values include the average (standard deviation) of performance metrics in Example 1. }\label{tab:npvar1}
    \begin{tabular}{lccccccc}
    \hline
     &&\multicolumn{3}{c}{$p$} &\multicolumn{3}{c}{$(n_1, n_2)$}\\
     \cmidrule(lr){3-5}\cmidrule(lr){6-8}
        Method & Metric & 50 & 100 & 150 & (50, 80) & (100, 200) & (200, 300) \\ \hline
        MultiCOAP & A\_tr & 0.99(4e-3)   & 0.99(1e-3)   & 0.98(4e-3)   & 0.97(4e-3)   & 0.99(1e-3)   & 0.99(7e-4) \\
         ~ & F\_tr & 0.92(0.01)   & 0.94(0.01)   & 0.95(0.01)   & 0.93(0.01)   & 0.95(0.01)   & 0.95(4e-3) \\
         ~ & beta\_er& 0.18(0.05)   & 0.12(0.01)   & 0.10(0.01)   & 0.14(0.02)   & 0.11(0.01)   & 0.10(0.01) \\
         ~ & B\_tr& 0.81(0.05)   & 0.84(0.03)   & 0.81(0.03)   & 0.75(0.03)   & 0.85(0.03)   & 0.91(0.02) \\
         ~ & H\_tr& 0.60(0.12)   & 0.75(0.05)   & 0.73(0.05)   & 0.71(0.05)   & 0.75(0.06)   & 0.79(0.05) \\
        \hline
         MSFR & A\_tr& 0.30(0.06)   & 0.46(0.07)   & 0.58(0.07)   & 0.36(0.10)   & 0.43(0.08)   & 0.45(0.08) \\
         ~ & F\_tr& 0.15(0.06)   & 0.29(0.10)   & 0.44(0.08)   & 0.17(0.06)   & 0.31(0.09)   & 0.41(0.07) \\
         ~ & beta\_er& 681(899)     & 15.6(16.8)   & 4.35(1.29)   & 31.5(30.6)   & 21.0(41.1)   & 11.0(4.00) \\
         ~ & B\_tr & 0.07(0.04)   & 0.10(0.05)   & 0.12(0.04)   & 0.09(0.05)   & 0.10(0.05)   & 0.11(0.04) \\
         ~ & H\_tr& 0.06(0.03)   & 0.09(0.05)   & 0.16(0.06)   & 0.11(0.05)   & 0.09(0.05)   & 0.09(0.05) \\
         \hline
         MSFR-log & A\_tr& 0.91(0.09)   & 0.95(0.07)   & 0.95(0.06)   & 0.90(0.08)   & 0.95(0.06)   & 0.95(0.07) \\
         ~ & F\_tr& 0.74(0.09)   & 0.82(0.08)   & 0.84(0.07)   & 0.69(0.09)   & 0.83(0.09)   & 0.85(0.11) \\
         ~ & beta\_er& 0.38(0.01)   & 0.32(0.01)   & 0.29(3e-3)   & 0.34(0.01)   & 0.31(4e-3)   & 0.31(4e-3) \\
         ~ & B\_tr& 0.71(0.06)   & 0.71(0.05)   & 0.70(0.04)   & 0.60(0.06)   & 0.72(0.05)   & 0.77(0.05) \\
         ~ & H\_tr& 0.49(0.06)   & 0.55(0.08)   & 0.61(0.06)   & 0.54(0.08)   & 0.56(0.08)   & 0.56(0.08) \\
        \hline
         COAP & A\_tr& 0.96(0.03)   & 0.94(0.03)   & 0.94(0.04)   & 0.91(0.04)   & 0.93(0.03)   & 0.96(0.02) \\
         ~ & F\_tr& 0.88(0.04)   & 0.87(0.03)   & 0.88(0.06)   & 0.81(0.05)   & 0.85(0.04)   & 0.89(0.02) \\
         ~ & beta\_er & 0.18(0.02)   & 0.13(0.01)   & 0.12(0.01)   & 0.16(0.02)   & 0.13(0.01)   & 0.11(0.01) \\
    \hline
    \end{tabular}
\end{table}

\underline{\bf Example 2}. In this illustrative scenario, we explore the impact of overdispersion, a common phenomenon in count data. Overdispersion often manifests in practical scenarios, prompting our investigation into its influence. We systematically escalate $\sigma_0^2$ values from $1$ to $4$ and subsequently to $8$. A larger $\sigma^2_0$ indicates higher overdispersion, enabling us to assess the effects of this phenomenon. Additionally, we set $p=100, (n_1, n_2)=(100,200), (\rho_A, \rho_B, \rho_z)=(2,3.5,1)$, maintaining other parameters consistent with Example 1. The findings in Table \ref{tab:overdisper} indicate that MultiCOAP, alongside other methods, experiences a decline in performance with increasing overdispersion strength. However, MultiCOAP consistently outperforms others, maintaining a substantial lead. Furthermore, applying MSFR directly to count data yields notably poor performance, with the algorithm breaking down as overdispersion reaches $4$ and $8$. Although the log-normalized data version of MSFR (MSRF-log) also shows a good performance for the estimation of the study-shared loadings but it performs much worse than MultiCOAP in the estimation of other parameters.
\begin{table}[h!]
    \centering\scriptsize
  \caption{Comparison of MultiCOAP and other competitors for parameter estimation.  Reported are the average (standard deviation) for performance metrics in Example 2. ``$-$" means that the algorithm of the corresponding method breaks down under the corresponding data setting.}\label{tab:overdisper}
    \begin{tabular}{lcccccccc}
    \hline
     Metric   & Method & $\sigma_0^2=1$ & $\sigma_0^2=4$ & $\sigma_0^2=8$ & Method  & $\sigma_0^2=1$ & $\sigma_0^2=4$ & $\sigma_0^2=8$ \\ \cmidrule(lr){1-5}\cmidrule(lr){6-9}
 A\_tr   & MultiCOAP & 0.99(1e-3) & 0.97(0.01)   & 0.92(0.03) & MSFR-log  & 0.96(0.06)   & 0.96(0.01)   & 0.91(0.02) \\
 F\_tr  && 0.95(0.01)   & 0.89(0.01)   & 0.79(0.04)  & & 0.84(0.08)   & 0.82(0.03)   & 0.72(0.04) \\
  beta\_er && 0.11(0.01)   & 0.19(0.02)   & 0.28(0.01) &  & 0.31(4e-3) & 0.39(3e-3) & 0.47(4e-3) \\
   B\_tr && 0.85(0.03)   & 0.68(0.04)   & 0.49(0.06)  & & 0.72(0.05)   & 0.61(0.06)   & 0.45(0.06) \\
   H\_tr && 0.75(0.06)   & 0.53(0.04)   & 0.33(0.05)  & & 0.56(0.07)   & 0.47(0.05)   & 0.30(0.04) \\
   \cmidrule(lr){1-5}\cmidrule(lr){6-9}
  A\_tr  & MSFR &  0.44(0.08)   & -            & -            & COAP & 0.93(0.03)   & 0.91(0.03)   & 0.89(0.04) \\
  F\_tr  &&  0.32(0.06)   & -            & -           & & 0.85(0.03)   & 0.80(0.04)   & 0.74(0.05) \\
  beta\_er &&  20.0(21.6)   & -            & -          &  & 0.13(0.01)   & 0.20(0.01)   & 0.27(0.01) \\
  B\_tr &&  0.10(0.04)   & -            & -           & & -            & -            & -         \\
  H\_tr &&  0.09(0.05)   & -            & -           & & -            & -            & -         \\
    \hline
    \end{tabular}
\end{table}

\underline{\bf Example 3}. Apart from the absolute values of count variables, practitioners often find significance in the relative values of counts, specifically the relative gene expression values in scRNA-seq data. Consequently, we generate relative counts by setting $a_{si}$ greater than 1. This is achieved by randomly drawing an integer for $a_{si}$ from an interval $[a, b]$, while keeping other parameters consistent with Example 2. Table \ref{tab:relative} presents the results, excluding MSFR due to its algorithm breaking down in all runs under this setting. Across all considered configurations, MultiCOAP consistently outperforms the other two methods. As the values of the normalized factor $a_{si}$ increase, all methods witness a decrease in parameter estimation accuracy. Notably, COAP experiences a substantial decline in estimation performance when the interval $[a, b]$ changes from $[41, 50]$ to $[101, 110]$. MSFR-log also exhibits poor estimation for $\bb$, approximately ten times higher in estimation error than that of MultiCOAP. In contrast, MultiCOAP demonstrates the highest level of robustness in these scenarios.

\begin{table}[h!]
    \centering\renewcommand\tabcolsep{3pt}
  \caption{ Comparison of MultiCOAP and other methods for parameter estimation.  Reported are the average (standard deviation) for performance metrics in Example 3 with  relative counts. ``$-$" means that the algorithm of the corresponding method breaks down under the corresponding data setting. Note that the algorithm of MSFR breaks down in all runs under this setting.}\label{tab:relative}
    \begin{tabular}{lcccccc}
    \hline
        Method & $[a, b]$ & A\_tr & F\_tr & beta\_er &  B\_tr & H\_tr  \\ \hline
  MultiCOAP & $[11,20]$ & 0.99(6e-4) & 0.95(0.01)   & 0.09(0.01)   & 0.90(0.01)   & 0.86(0.02)  \\
   ~ & $[41,50]$ & 0.99(5e-4) & 0.95(0.01)   & 0.10(0.02)   & 0.91(0.01)   & 0.87(0.01)  \\
   ~ & $[101,110]$ & 0.99(0.01)   & 0.93(0.02)   & 0.10(0.01)   & 0.86(0.06)   & 0.85(0.07)  \\\hline
   MSFR-log  & $[11,20]$ & 0.99(1e-3) & 0.91(0.01)   & 0.92(0.01)   & 0.81(0.04)   & 0.77(0.06)  \\
 ~  & $[41,50]$ & 0.99(0.01)   & 0.91(0.04)   & 1.23(5e-3) & 0.84(0.04)   & 0.85(0.05)  \\
 ~ & $[101,110]$ &  0.97(0.05)   & 0.88(0.10)   & 1.49(5e-3) & 0.83(0.07)   & 0.82(0.13)  \\\hline
 COAP  & $[11,20]$ & 0.97(0.01)   & 0.92(0.02)   & 0.10(0.02)   & -          & -        \\
 ~ & $[41,50]$ &  0.93(0.02)   & 0.92(0.02)   & 0.10(0.02)   & -           & -         \\
  ~ & $[101,110]$ & 0.83(0.03)   & 0.85(0.03)   & 0.10(0.02)   & -          & -         \\
    \hline
    \end{tabular}
\end{table}

\underline{\bf Example 4}.  In this example, we explore the impact of signal strength balance between study-shared factors and study-specified factors by considering different combinations of $(\rho_A, \rho_B)$, specifically, $(0.8, 1)$, $(2, 1)$, and $(2, 3)$. We maintain $\rho_z=1$, ${(n_1, n_2)=(150, 200)}, p=100, d=r=3$, and keep other parameters consistent with Example 1. The findings in Table \ref{tab:balance} indicate that increasing the signal strength of the study-shared factor, denoted by $\rho_B$ rising from $0.8$ to $2$, not only enhances the estimation accuracy of the study-shared factors and loadings but also improves the estimation performance of study-specified factors and loadings. This improvement stems from the fact that better performance in estimating study-shared factors and loadings positively influences the estimation of study-specified parameters. As the signal strength of factors increases, the relative signal in $\z_{si}$ decreases, leading to a slight increase in the estimation error of $\bb$. Simultaneously, we observe that increasing the signal strength of the study-specified factor, denoted by $\rho_A$ rising from $1$ to $3$, solely results in improved estimation accuracy of study-specified factors and loadings.

\begin{table}[h!]
    \centering\caption{Comparison of MultiCOAP and other competing models for parameter estimation. The reported values include the average (standard deviation) of performance metrics in Example 4.}\label{tab:balance}
    \begin{tabular}{lcccc}
    \hline
        Method &$(\rho_A, \rho_B)$ & $(0.8, 1)$ & $(2, 1)$ & $(2, 3)$ \\ \hline
        MultiCOAP & A\_tr & 0.93(0.02) & 0.99(1e-3) & 0.99(1e-3) \\
        ~ & F\_tr & 0.80(0.04) & 0.96(0.01) & 0.95(0.01) \\
        ~ & beta\_er & 0.16(0.01) & 0.19(0.02) & 0.18(0.01) \\
        ~ & B\_tr & 0.43(0.06) & 0.68(0.07) & 0.87(0.02) \\
        ~ & H\_tr & 0.25(0.04) & 0.49(0.07) & 0.77(0.05) \\ \hline
        MSFR-log & A\_tr & 0.92(0.02) & 0.83(0.09) & 0.89(0.10) \\
        ~ & F\_tr & 0.75(0.03) & 0.70(0.09) & 0.77(0.13) \\
        ~ & beta\_er & 0.44(3e-3) & 0.54(0.01) & 0.55(0.01) \\
        ~ & B\_tr & 0.40(0.06) & 0.38(0.05) & 0.72(0.04) \\
        ~ & H\_tr & 0.24(0.04) & 0.24(0.04) & 0.48(0.06) \\ \hline
        MSFR & A\_tr & 0.83(0.05) & 0.49(0.08) & 0.49(0.07) \\
        ~ & F\_tr & 0.63(0.04) & 0.47(0.09) & 0.42(0.08) \\
        ~ & beta\_er & 1.20(0.03) & 14.48(6.58) & 23.04(25.65) \\
        ~ & B\_tr & 0.15(0.08) & 0.04(0.03) & 0.08(0.04) \\
        ~ & H\_tr & 0.09(0.05) & 0.02(0.01) & 0.08(0.06) \\ \hline
        COAP & A\_tr & 0.94(0.01) & 0.99(1e-3) & 0.98(0.01) \\
        ~ & F\_tr & 0.80(0.01) & 0.96(5e-3) & 0.94(0.01) \\
        ~ & beta\_er & 0.16(5e-3) & 0.18(0.01) & 0.20(0.01) \\ \hline
    \end{tabular}
\end{table}

\underline{\bf Example 5}. In this example, we consider the selection accuracy of  the number of factors and conduct the sensitivity analysis of errorous selection of factors. We select the number of factors using the method CUP introduced in Section \ref{sec:select}.  We set $(\rho_A, \rho_B, \rho_z)=(2,5,1)$ while other parameters as same as Example 4. To assess the performance of our proposed methods, we conduct comparisons with the singular value ratio-based (SVR) method as outlined in \cite{liu2024highdimensional} for MultiCOAP, the Akaike information criterion (AIC), and the Bayesian information criterion (BIC) utilized in MSFR~\citep{de2023multi}. Additionally, we consider variants adapted for log-normalized data, denoted as BIC-log and AIC-log. The criterion value is calculated on a grid of $q\in \{1, \cdots, 6\}$ and $q_s \in \{1, \cdots, 4\}$. Correspondingly, we set $\tau=95\%, q_{max}=6$ and $q_{s,max}=4$ used in our proposed method and SVR method in \cite{liu2024highdimensional}.
Table \ref{tab:select} illustrates that the proposed CUP criterion outperforms other methods significantly across various noise strengths, particularly for the selection of $q$. However, with an increase in noise strength, the effectiveness of CUP in selecting $q_s$ notably diminishes. This observation is reasonable, given the inadequacy of signal in the analyzed data. Consequently, in the subsequent analysis, we delve into the performance of the estimators when the number of factors is misspecified under insufficient signal conditions.
\begin{table}[h!]
\centering\caption{Comparison of the proposed 'CUP' method  and other existing methods in model selection performance. Reported are the average of estimated number of factors in Example 5, where SD denotes standard deviation.}\label{tab:select}
\begin{tabular}{lrrrrrrr}
  \hline
 Case && CUP & SVR & AIC-log & BIC-log & AIC & BIC \\
  \hline
$\sigma^2=1$&$q=3$ & 3.00 & 4.09 & 6.00 & 4.09 & 3.26 & 3.28 \\
 & SD & 0.00 & 0.67 & 0.00 & 0.29 & 2.19 & 1.85 \\
  &$q_1=2$ & 2.14 & 1.78 & 2.84 & 1.00 & 3.25 & 2.59 \\
  &SD &  0.40 & 0.88 & 0.63 & 0.00 & 1.18 & 1.12 \\
   &$q_2=2$ & 2.00 & 1.21 & 2.84 & 1.00 & 3.25 & 2.59 \\
  &SD &  0.00 & 0.52 & 0.63 & 0.00 & 1.18 & 1.12 \\
   \hline
$\sigma^2=2$  &$q=3$ & 3.00 & 3.51 & 5.54 & 2.99 & 1.48 & 1.64 \\
&  SD & 0.00 & 0.67 & 0.59 & 0.17 & 1.29 & 1.04 \\
&  $q_1=2$ &  2.91 & 1.50 & 1.15 & 1.00 & 2.36 & 1.68 \\
&  SD & 0.29 & 0.77 & 0.36 & 0.00 & 1.25 & 0.75 \\
&  $q_2=2$ &  2.85 & 1.03 & 1.15 & 1.00 & 2.36 & 1.68 \\
&  SD & 0.36 & 0.22 & 0.36 & 0.00 & 1.25 & 0.75 \\
   \hline
\end{tabular}
\end{table}
We consider two cases to generate data. In Case 1, we set the true $q_s$ and misspecified $q=2, 3$ or $5$ in MultiCOAP and other compared methods, to evaluate the influence of misselected $q$. In Case 2, we set the true $q$ and misspecified $q_s=1, 2$, or $4$ to evaluate the influence of misselected $q_s$. Note that the true value of $q$ is $3$ and $q_s$ is $2$. In case 1, we only compare MultiCOAP with MSRF and MSFR-log since only these methods involve the selection of $q_s$. Table \ref{tab:misq} reveals that the under-selection of $q$ significantly impacts the estimation accuracy of both study-shared and study-specified factors and loadings. Conversely, the over-selection of $q$ does not exert a significant effect on the study-shared factors and loadings but does have a considerable impact on the study-specified factors and loadings of MultiCOAP. In general, MultiCOAP outperforms other competitors, except for the estimation performance of study-specified factors and loadings under the over-selection of $q$. When $q$ is overestimated, the study-specified factors become partially projected onto the study-shared factors, leading to a decrease in the estimation performance of the study-specified quantities. Moreover, we observe that the misselection of the number of study-specified factors ($q_s$) does not influence the estimation accuracy of study-shared factors and loadings but has a significant effect on the estimation accuracy of study-specified factors and loadings. Fortunately, the over-selection of
$q_s$  does not have a negative impact on both study-specified/shared factors and loadings. Thus, we can opt for more study-specified factors based on our proposed methods for a robust application.

\begin{table}[h!]
    \centering\renewcommand\tabcolsep{3pt}
    \caption{Comparison of MultiCOAP and other competing models for parameter estimation. The reported are the average (standard deviation) of performance metrics in Example 5 with misspecified number of factors.}\label{tab:misq}
    \begin{tabular}{lccccccc}
    \hline
        Method & Metric & $q=2$ & $q=3$ & $q=5$ & $q_s=1$ & $q_s=2$ & $q_s=4$ \\ \hline
        ~ & A\_tr & 0.79(0.01) & 0.96(0.02) & 0.96(0.01) & 0.96(0.01) & 0.96(0.02) & 0.96(0.02) \\
        ~ & F\_tr & 0.60(0.02) & 0.85(0.02) & 0.87(0.01) & 0.85(0.02) & 0.85(0.02) & 0.85(0.04) \\
        MultiCOAP & beta\_er & 0.17(0.01) & 0.17(0.01) & 0.17(0.01) & 0.17(0.01) & 0.17(0.01) & 0.17(0.01) \\
        ~ & B\_tr & 0.60(0.05) & 0.66(0.05) & 0.18(0.08) & 0.56(0.07) & 0.66(0.05) & 0.69(0.04) \\
        ~ & H\_tr & 0.36(0.04) & 0.51(0.03) & 0.21(0.05) & 0.34(0.04) & 0.51(0.03) & 0.53(0.03) \\ \hline
        MSFR-log & A\_tr & 0.69(0.07) & 0.94(0.01) & 0.94(0.01) & 0.94(0.01) & 0.94(0.01) & 0.89(0.06) \\
        ~ & F\_tr & 0.57(0.02) & 0.83(0.02) & 0.83(0.02) & 0.85(0.01) & 0.83(0.02) & 0.72(0.08) \\
        ~ & beta\_er & 0.48(3e-3) & 0.48(3e-3) & 0.48(3e-3) & 0.48(3e-3) & 0.48(3e-3) & 0.48(3e-3) \\
        ~ & B\_tr & 0.57(0.05) & 0.64(0.05) & 0.62(0.05) & 0.55(0.05) & 0.64(0.05) & 0.66(0.04) \\
        ~ & H\_tr & 0.36(0.04) & 0.49(0.04) & 0.48(0.04) & 0.35(0.04) & 0.49(0.04) & 0.51(0.03) \\ \hline
        MSFR & A\_tr & 0.59(0.07) & 0.73(0.05) & 0.77(0.03) & 0.72(0.06) & 0.73(0.05) & 0.71(0.06) \\
        ~ & F\_tr & 0.48(0.03) & 0.65(0.05) & 0.67(0.03) & 0.65(0.06) & 0.65(0.05) & 0.60(0.06) \\
        ~ & beta\_er & 1.69(0.07) & 1.70(0.08) & 1.80(0.19) & 1.73(0.11) & 1.70(0.08) & 1.69(0.12) \\
        ~ & B\_tr & 0.26(0.06) & 0.28(0.07) & 0.27(0.08) & 0.19(0.09) & 0.28(0.07) & 0.36(0.06) \\
        ~ & H\_tr & 0.19(0.04) & 0.20(0.05) & 0.20(0.06) & 0.15(0.07) & 0.20(0.05) & 0.27(0.04) \\ \hline
        COAP & A\_tr & 0.78(0.01) & 0.88(0.03) & 0.96(4e-3) & ~ & ~ & ~ \\
        ~ & F\_tr & 0.60(0.02) & 0.76(0.04) & 0.86(0.01) & ~ & ~ & ~ \\
        ~ & beta\_er & 0.18(0.01) & 0.17(0.01) & 0.16(0.01) \\ \hline
    \end{tabular}
\end{table}

\underline{\bf Example 6}. In practical applications, particularly in genomics, both sample size and count variable dimensions are often very high. Consequently, an effective method should exhibit rapid computational efficiency. To investigate this efficiency, we compare MultiCOAP with other competitors in terms of running time, progressively increasing both sample size and count variable dimension. We consider two cases:
In Case 1, we maintain $p=800$ while varying the sample size within the set  $\{(n_1, n_2)=c \times (200, 300): c=1, 4, \cdots, 16, 19\}$. In Case 2, we fix $(n_1, n_2)=(1000, 2000)$ while altering the count variable dimension within $\{200, 500, 800, 900, 2000, 5000, 8000, 1e4\}$. Table \ref{tab:time} presents the average running time of MultiCOAP and other methods over ten repeats. The findings suggest that, despite MultiCOAP incorporating the estimation of study-specified factors, it remains comparable to COAP in terms of computational efficiency. Additionally, among the methods that incorporate study-specified factors, MultiCOAP exhibits markedly superior computational efficiency compared to both MSFR-log and MSFR. Notably, MSFR demonstrates instability when dealing with count data characterized by high dimensions in sample size or variables directly, while MSFR-log encounters challenges in handling data with high-dimensional variables.

\begin{table}[h!]
    \centering\scriptsize 
    \caption{Comparison of computational time (in second) for MultiCOAP and other methods. (a) The
running time versus total sample size $n$ in case 1 under $p=800$; (b) the running time versus variable dimension
$p$ in case 2 under $(n_1, n_2)=(1000, 2000)$. "-" means the algorithm of the corresponding method breaks down under the setting.}\label{tab:time}
    \begin{tabular}{llllllll}
    \hline
         & \multicolumn{7}{c}{$n$} \\
\cmidrule(lr){2-8}
       Method &  500 & 2000 & 3500 & 5000 & 6500 & 8000 & 9500 \\ \hline
        MultiCOAP & 1.67(0.04) & 6.82(0.10) & 13.15(0.60) & 18.24(1.13) & 22.54(0.07) & 28.03(0.43) & 36.37(3.43) \\
        COAP & 1.62(0.04) & 6.92(0.10) & 13.45(0.70) & 18.83(1.24) & 25.59(0.97) & 32.00(1.53) & 37.87(1.49) \\
        MSFRlog & 1702(10.0) & 1784(10.0) & 1993(17.0) & 2052(18.0) & 2087(22.0) & 2165(177) & 2202(255) \\
        MSFR & - & - & - & - & - & - & - \\ \hline
         & \multicolumn{7}{c}{$p$} \\
\cmidrule(lr){2-8}
         & 200 & 500 & 800 & 900 & 2000 & 5000 & 8000 \\ \hline
        MultiCOAP & 2.44(0.04) & 6.45(0.10) & 10.44(0.15) & 12.21(0.52) & 27.28(0.47) & 82.56(9.33) & 132(14.06) \\
        COAP & 2.52(0.05) & 6.56(0.10) & 10.63(0.14) & 12.55(0.53) & 27.96(0.80) & 74.54(4.85) & 119(8.07) \\
        MSFRlog & 140(0.58) & 722(6.90) & 1808(43.8) & 2623(267.2) & - & - & - \\
        MSFR & 138(1.51) & - & - & - & - & - & - \\ \hline
    \end{tabular}
\end{table}
\nvs
\section{Real data analysis}\label{sec:real}
To demonstrate the efficacy of MultiCOAP, we employed it in the analysis of the  motivating PBMC data obtained through CITE-seq sequencing technology. This data was derived from a case-control study, wherein the case group underwent stimulation, as detailed in the work by \cite{mimitou2021scalable}. For each group, CITE-seq concurrently measures the expression levels of 33,538 genes and 227 protein markers across thousands of cells, with 3,480 and 4,645 cells in case and control groups, respectively. Following the preprocessing method for CITE-seq sequencing data  in \cite{stoeckius2017simultaneous}, we initially selected the top 2,000 highly variable genes with high quality. Our primary objective was to investigate the homogeneity and heterogeneity of gene expression patterns between the case and control groups, and the associations  between genes and proteins. This involved extracting study-shared factors and study-specified factors while accounting for the impact of protein markers on gene expressions.

Before applying MultiCOAP to this data, we employed the proposed CUP method to determine tuning parameters by setting $(q_{\text{max}}, q_{\text{s,max}}, r_{\text{max}}) =(20, 10, 10)$. This resulted in $(\hat{q}, \hat{q}_1, \hat{q}_2, \hat{r})=(6, 9, 3, 3)$, with $\tau=95\%$ chosen to explain more variations in the data. Subsequently, we fitted MultiCOAP using these selected values.
To demonstrate its superior performance, we also applied COAP, MSFR, and MSFR-log, using the same number of factors and rank employed in MultiCOAP for a fair comparison. However, due to the high dimensionality ($p=2,000$ and $d=227$), both MSFR and MSFR-log experienced breakdown, consistent with the observed instability in the simulation study. Therefore, we solely benchmarked MultiCOAP against COAP.

As there are no true values of parameters in real data, we adopt an index called adjusted MacFadden's $\mathrm{R}^2$~\citep{mcfadden1987regression} to measure the feature extraction performance. For the $s$-th study, we denote the extracted feature matrix as $\wh{\bm{R}}_s$ and $\x_{s.j}$ represents the $j$-th column of $\X_s$. For each gene $j$, we fitted a Poisson regression model between $\x_{s.j}$ and $\wh{\bm{R}}_s$, and calculated the adjusted MacFadden's $\mathrm{R}^2$ based on the fitted model. This metric measures how much information of $\x_{s.j}$ is contained in $\wh{\bm{R}}_s$, and a larger value implies better performance in feature extraction. For COAP, the feature matrix is defined as $\wh{\bm{R}}_{s}^{coap}=(\wh \H^{coap}_s, \Z_s \wh \V_{\bb}^{coap})$, where $\wh \H^{coap}_s$ is the estimate of latent factor matrix of study $s$ and $\wh \V_{\bb}^{coap}\in \mathbb{R}^{d\times \widehat r}$ is the rank-$\widehat r$ right singular matrix of $\wh\bb^{coap}$. For MultiCOAP, we define the feature matrix as $\wh{\bm{R}}_{s1}=(\wh\F_s, \wh \H_s, \Z_s \wh \V_{\bb})$, where $\wh\H_s$ is the estimate of the study-specified factor matrix for the study $s$ not considered in COAP, $\wh\F_s$ is the estimate of the study-shared factor matrix, and $\wh \V_{\bb}\in \mathbb{R}^{d\times \widehat r}$ is the rank-$\widehat r$ right singular matrix of $\wh\bb$. We also consider another definition $\wh{\bm{R}}_{s2}=(\wh\F_s, \Z_s \wh \V_{\bb})$ not using $\wh \H_s$, to fairly compare with COAP. We divided the genes into ten bins, with 200 genes in each bin, and calculated the mean of adjusted MacFadden's $\mathrm{R}^2$ within each bin. As illustrated in Figure \ref{fig:R2}, MultiCOAP consistently demonstrated superior performance over COAP in feature extraction across all ten bins and two studies. Remarkably, even when removing the study-specified factor matrix from the features, MultiCOAP continued to outperform COAP. This emphasizes that accounting for underlying heterogeneity among studies using the study-specified factors contributes to the improved estimation of the study-shared factor components.
\begin{figure}[h!]
  \centering
  \includegraphics[width=0.9\textwidth ]{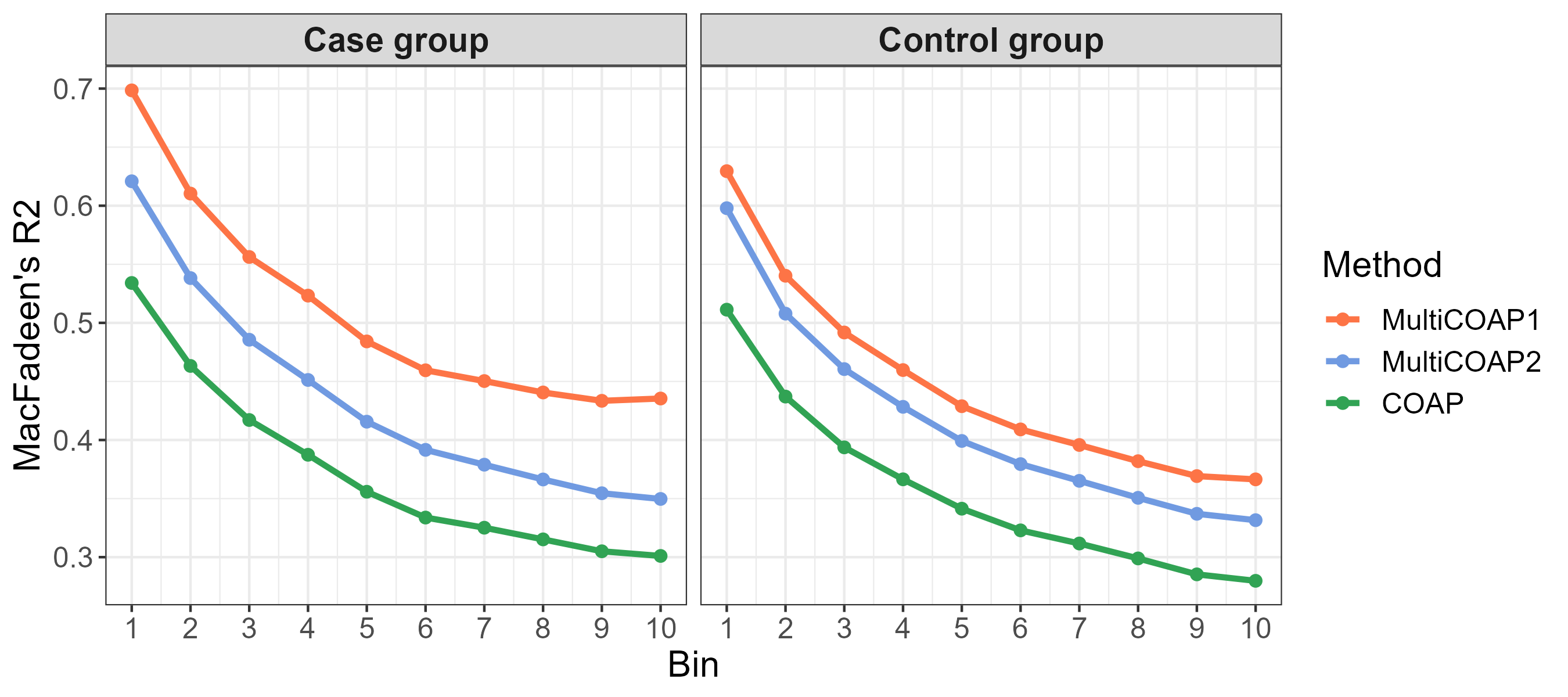}
  \caption{Evaluation of biological feature extraction performance comparing MultiCOAP and COAP based on adjusted MacFadden's $\mathrm{R}^2$. MultiCOAP1 and MultiCOAP2 denote the utilization of $\wh{\bm{R}}_{s1}$  and $\wh{\bm{R}}_{s2}$  to compute the adjusted MacFadden's $\mathrm{R}^2$, respectively.}\label{fig:R2}
\end{figure}

The extracted features obtained from MultiCOAP   provide valuable benefits in heterogeneous cell type identification. Based on the extracted feature matrix $\wh{\bm{R}}_{s1}$ from MultiCOAP, we perform the Louvain clustering to identify the heterogenous cell clusters in each study. We identified 10 cell clusters in each of case and control groups. We visualize these cell clusters on the two-dimensional UMAP components~\citep{mcinnes2018umap} in Figure \ref{fig:celltype}(a), which indicates the good separation among cell clusters in case and control groups. To determine the cell types, we perform the differential expression analysis among the identified cell clusters using the {\it FindAllMarkers} function in the {\it Seruat} R package, a common-used single-cell analysis toolbox, and identified 1743 and 1373 marker genes in total for case and control group, respectively, by setting the cutoff of the adjusted p-value less than 0.01 and the log-fold change less than 0.5. Supplementary Figure S2 showed that the marker gene expression patterns are also differentiable for the identified cell clusters.  As benchmark, we perform the same analysis using the extracted feature matrix $\wh{\bm{R}}_{s}^{coap}$ from COAP, while we identified 10 and eight cell clusters for case and control groups, respectively, with only 841 and 928 marker genes identified in total, respectively. This finding indicates that MultiCOAP enhanced the identification of marker genes by capturing more heterogenous information in data with usage of the study-specified factors. In the following, we annotate the cell types for the cell clusters identified by MultiCOAP using the identified marker genes and cell-type database, PanglaoDB (\url{https://panglaodb.se/}). The corresponding cell types for case and control groups are summarized in Figure \ref{fig:celltype}(b).  We observed there exist not only same cell types shared by  case and control group, such as Monocyte, NK cell, Dentritic cells and IGHM+ B cells, but also group-specified cell types, such as IL3RA+ B cells and CD84+ T cells in case group and GBP4+ T cell and CD69+ T cells in control group. These findings suggest that the model structure incorporating both shared and study-specific factors is appropriate.  Figure \ref{fig:celltype}(c) shows the expression pattern of marker genes of the two subtypes of B cells in the case group, which confirms the correctness of cell typing. The primary function of B cells is to participate in the adaptive immune response by producing antibodies~\citep{mauri2012immune}. To investigate the functions of two subtypes of B cells, we perform gene set enrichment analysis based on their identified marker genes and Gene Ontology (GO) database. Supplementary Figure S3 showed that IGHM+ B cells are involved in biological processes related to B cell differentiation and the negative regulation of B cell proliferation and exhibit molecular functions in immunoglobulin binding, while IL3RA+ B cells play a role in actin filament organization, the regulation of angiogenesis, and possess molecular functions such as protein kinase C binding and actin filament binding.
\begin{figure}[H]
  \centering
  \includegraphics[width=0.9\textwidth ]{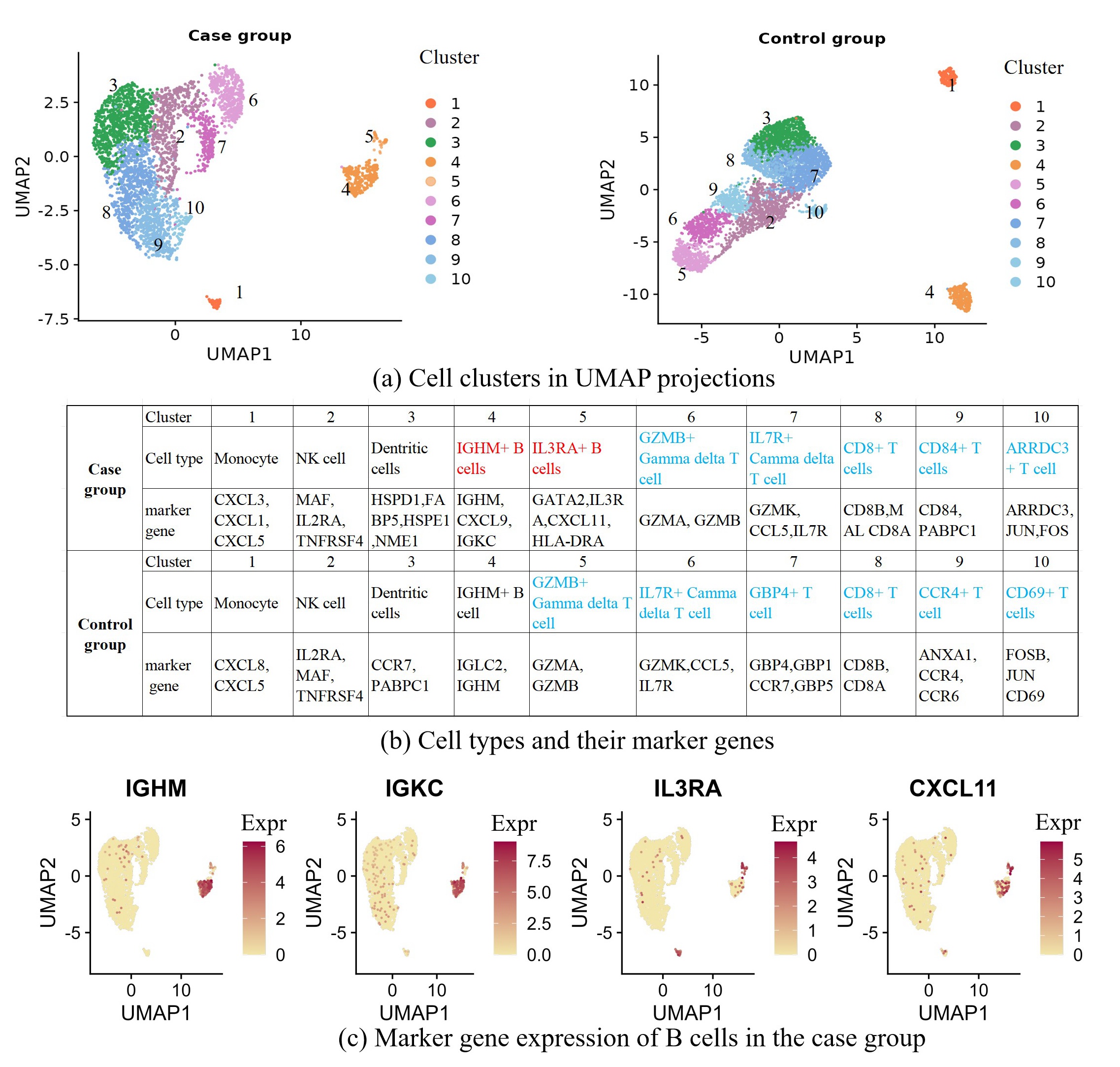}
  \caption{Cell clusters identified by MultiCOAP along with their corresponding cell types and marker genes. (a) Scatter plots displaying two-dimensional UMAP projections of the identified cell clusters; (b) Identification of cell types based on distinctive marker genes; (c) Scatter plots of two-dimensional UMAP projections for the expression levels of two markers genes corresponding to each of the two subtypes of B cells.}\label{fig:celltype}
\end{figure}
Unlike its competitors, MultiCOAP boasts unique features that enable the provision of uncertainty measures for the posterior estimator of each factor. Figure \ref{fig:factorband}(a) displays the sorted curves and confidence bands for each factor estimate. Notably, the first study-shared factor deviates significantly below zero in both studies, setting it apart from other factors that remain close to zero. To enhance interpretability, we visualize the value of this factor estimate in the UMAP plot and observe a concentration of signals on the Monocyte cells that play a multifaceted role in the immune system including essential components of the innate immune response and key players in coordinating immune defenses against various threats~\citep{evers2022single}. This observation suggests that the first study-shared factor effectively captures information about Monocyte cells, contributing to the interpretability of the factors.

\begin{figure}[h!]
  \centering
  \includegraphics[width=0.9\textwidth ]{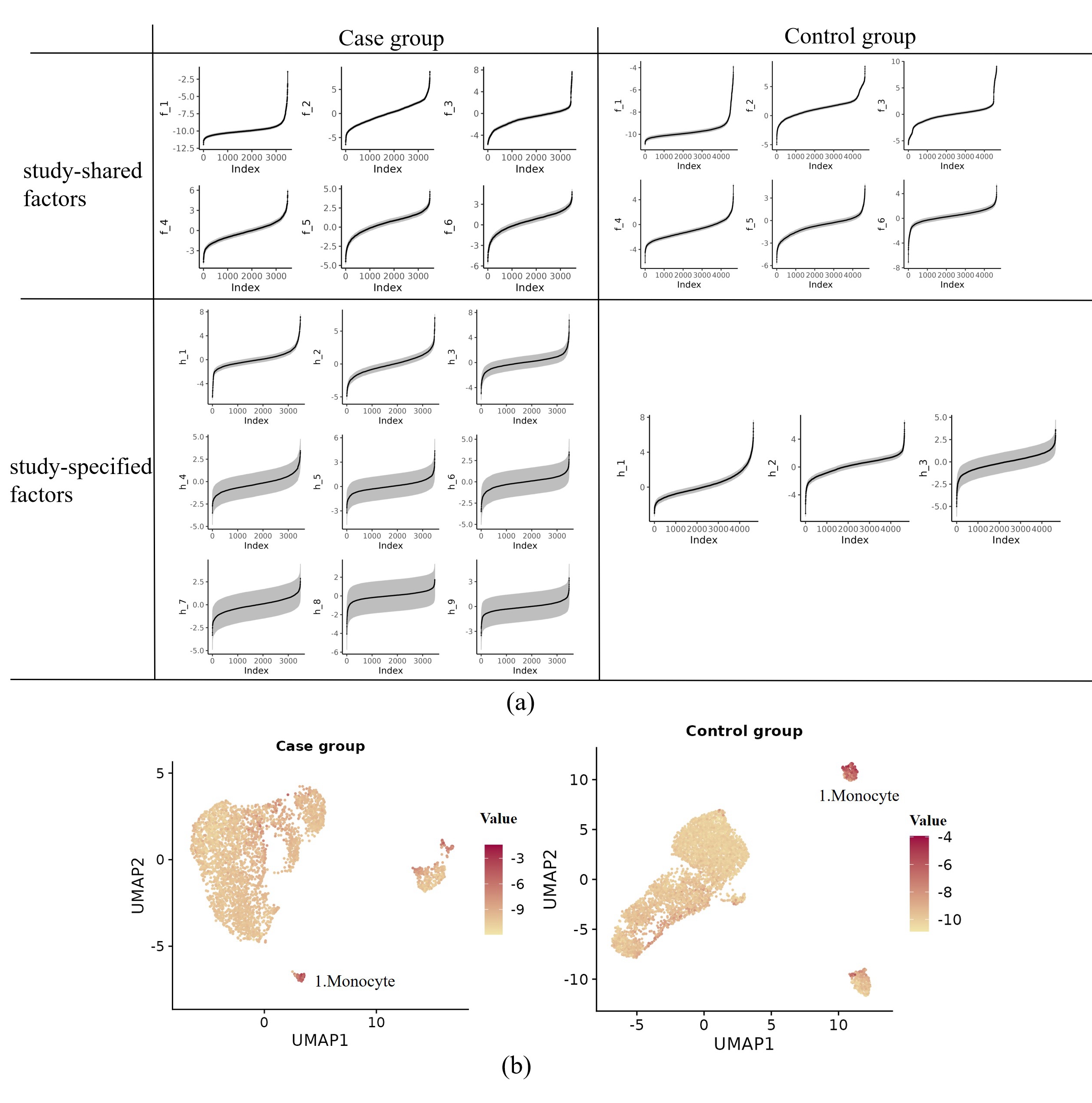}
  \caption{MultiCOAP benefits the cell type identification}\label{fig:factorband}
\end{figure}

Let $\tilde\bb$ denote the last 227 columns of $\wh\bb$, the estimator of the regression matrix $\wh\bb$ from MultiCOAP. Consequently,  $\tilde\bb\in \mathbb{R}^{2000 \times 227}$ captures the association between the expression levels of  2,000 genes and 227 proteins while excluding the influence of latent factors. For a given gene $j$,  $\tilde\bb_{j.}=(\tilde\beta_{j1}, \cdots, \tilde\beta_{j,227})$, where $\tilde{\beta}_{jk}$ represents the association strength between gene $j$ and protein $k$, with the entry having the largest value indicating the most strong association.
To investigate the association between genes and proteins in expression patterns, we identified the most associated proteins for marker genes of IGHM+ B cells, including IGHM, CXCL9, and IGKC (Figure \ref{fig:celltype}(b)). The top five associated proteins for each of these three marker genes consistently included CD19, CD25, and CD44. We visualized IGHM+ B cells alongside the expression levels of CD19 and CD25 on 2-dimensional UMAP projections of protein expressions as illustrated in Figure \ref{fig:proteinexp}(a). Our observations revealed a notably high expression pattern of both CD19 and CD25 on IGHM+ B cells in both case and control groups. These findings align with existing studies, indicating that CD19 is a cell surface protein predominantly expressed on B cells throughout their maturation process~\citep{depoil2008cd19}. Additionally, it is reported that CD25 is expressed in a subset of B cells and can be upregulated upon activation~\citep{brisslert2006phenotypic}. Figure \ref{fig:proteinexp}(b) indicates a dominant high expression level of CD25 in the case group compared to the control group, suggesting a potential activation response of B cells to the stimulation in the case group.

\begin{figure}[h!]
  \centering
  \includegraphics[width=0.9\textwidth ]{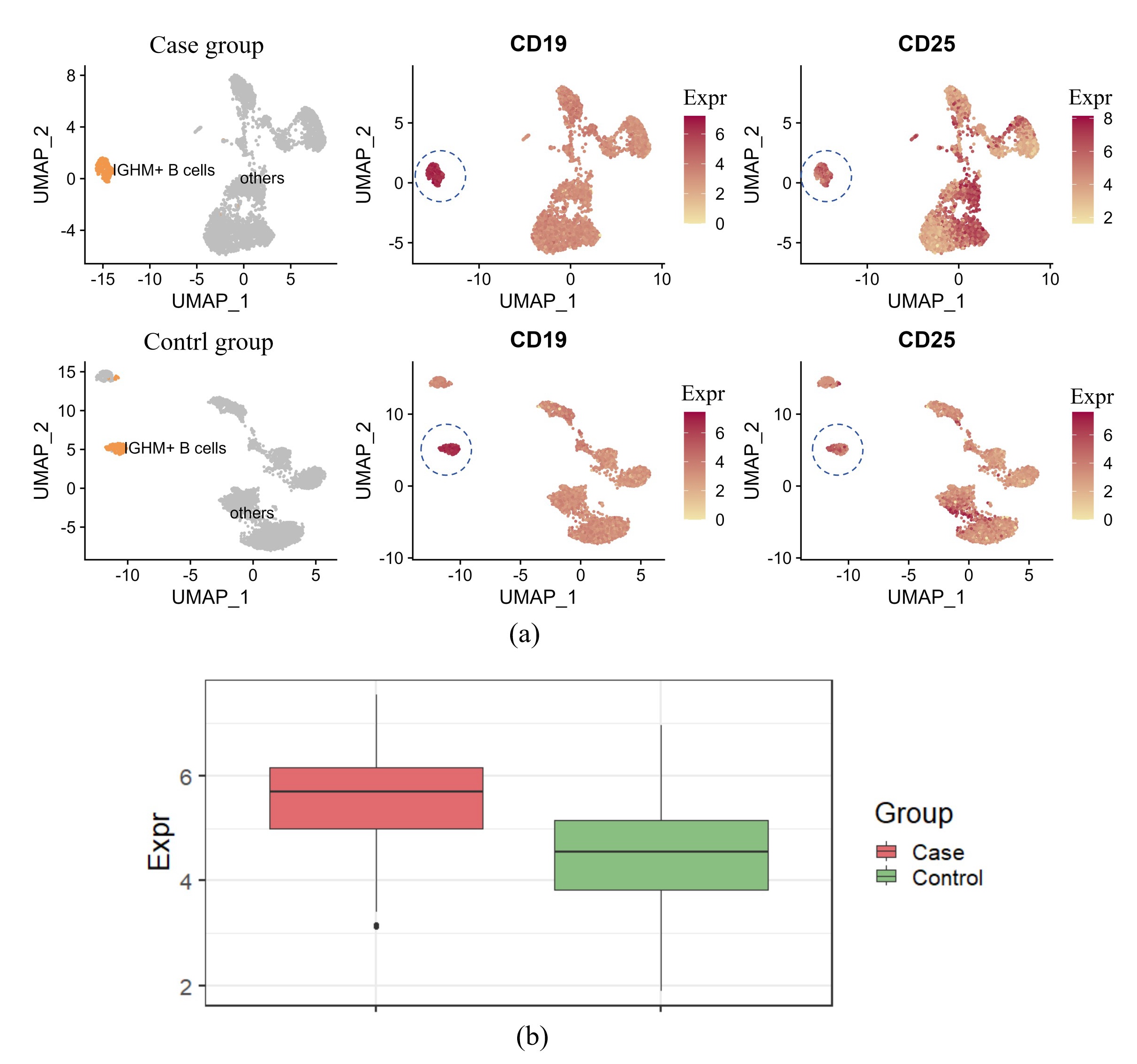}
  \caption{(a): Scattor plots of IGHM+ B cells alongside the expression levels of proteins CD19 and CD25 on 2-dimensional UMAP projections of protein expressions; (b) Boxplot of the expression levels of protein CD25 within IGHM+ B cells in both case and control groups.}\label{fig:proteinexp}
\end{figure}

\nvs
\section{Discussion}\label{sec:dis}
In this study, we introduce a novel model: the multi-study covariate-augmented overdispersed Poisson factor model. This model is tailored for effectively integrating high-dimensional count data from multiple studies or sources with overdispersion and additional covariates, by accounting for the shared homogeneity across studies and the inherent heterogeneity among them. We delve into the theoretical underpinnings by examining model identifiability conditions and employ variational likelihood estimation to facilitate model estimation, particularly in managing three latent large random matrices. Through this framework, we establish the convergence rates and asymptotic normality of estimators for model parameters, relaxing existing constraints on count variable dimensions. Furthermore, we devise a computationally efficient variational EM algorithm for practical implementation, and  a procedure for determining the optimal number of latent factors and the rank of the regression coefficient matrix.  Extensive numerical experiments demonstrate that MultiCOAP surpasses state-of-the-art methods in terms of model parameter estimation and computational efficiency, thereby facilitating more effective and promising real data analysis.


Moving forward, it is interesting to extend MultiCOAP to integrate multimodality data from multiple studies that have distinct variable type in each modality based on the framework of generalized factor models~\citep{liu2023generalized}.
Moreover, within the MultiCOAP framework, an emerging and exciting area of research is the exploration of spatial genomics data from diverse studies~\citep{liu2023probabilistic}. This direction not only encompasses the expression data of multi-omics but also incorporates the measured spatial location for each cell. This integration allows us to quantitatively assess the spatial dependencies among cells, offering deeper insights into the spatial organization and interactions within cellular systems.

\nvs
\bibliographystyle{Chicago}
\bibliography{ref}

\begin{thebibliography}{}

\bibitem[\protect\citeauthoryear{Argelaguet, Arnol, Bredikhin, Deloro, Velten,
  Marioni, and Stegle}{Argelaguet et~al.}{2020}]{argelaguet2020mofa+}
Argelaguet, R., D.~Arnol, D.~Bredikhin, Y.~Deloro, B.~Velten, J.~C. Marioni,
  and O.~Stegle (2020).
\newblock Mofa+: a statistical framework for comprehensive integration of
  multi-modal single-cell data.
\newblock {\em Genome biology\/}~{\em 21\/}(1), 1--17.

\bibitem[\protect\citeauthoryear{Bickel, Choi, Chang, and Zhang}{Bickel
  et~al.}{2013}]{peterAsy}
Bickel, P., D.~Choi, X.~Chang, and H.~Zhang (2013).
\newblock {Asymptotic normality of maximum likelihood and its variational
  approximation for stochastic blockmodels}.
\newblock {\em The Annals of Statistics\/}~{\em 41\/}(4), 1922 -- 1943.

\bibitem[\protect\citeauthoryear{Brisslert, Bokarewa, Larsson, Wing, Collins,
  and Tarkowski}{Brisslert et~al.}{2006}]{brisslert2006phenotypic}
Brisslert, M., M.~Bokarewa, P.~Larsson, K.~Wing, L.~V. Collins, and
  A.~Tarkowski (2006).
\newblock Phenotypic and functional characterization of human cd25+ b cells.
\newblock {\em Immunology\/}~{\em 117\/}(4), 548--557.

\bibitem[\protect\citeauthoryear{Chiquet, Mariadassou, and Robin}{Chiquet
  et~al.}{2018}]{chiquet2018variational}
Chiquet, J., M.~Mariadassou, and S.~Robin (2018).
\newblock Variational inference for probabilistic poisson pca.
\newblock {\em The Annals of Applied Statistics\/}~{\em 12\/}(4), 2674--2698.

\bibitem[\protect\citeauthoryear{De~Vito and Avalos-Pacheco}{De~Vito and
  Avalos-Pacheco}{2023}]{de2023multi}
De~Vito, R. and A.~Avalos-Pacheco (2023).
\newblock Multi-study factor regression model: an application in nutritional
  epidemiology.
\newblock {\em arXiv preprint arXiv:2304.13077\/}.

\bibitem[\protect\citeauthoryear{De~Vito, Bellio, Trippa, and
  Parmigiani}{De~Vito et~al.}{2019}]{de2019multi}
De~Vito, R., R.~Bellio, L.~Trippa, and G.~Parmigiani (2019).
\newblock Multi-study factor analysis.
\newblock {\em Biometrics\/}~{\em 75\/}(1), 337--346.

\bibitem[\protect\citeauthoryear{De~Vito, Bellio, Trippa, and
  Parmigiani}{De~Vito et~al.}{2021}]{de2021bayesian}
De~Vito, R., R.~Bellio, L.~Trippa, and G.~Parmigiani (2021).
\newblock Bayesian multistudy factor analysis for high-throughput biological
  data.
\newblock {\em The annals of applied statistics\/}~{\em 15\/}(4), 1723--1741.

\bibitem[\protect\citeauthoryear{Depoil, Fleire, Treanor, Weber, Harwood,
  Marchbank, Tybulewicz, and Batista}{Depoil et~al.}{2008}]{depoil2008cd19}
Depoil, D., S.~Fleire, B.~L. Treanor, M.~Weber, N.~E. Harwood, K.~L. Marchbank,
  V.~L. Tybulewicz, and F.~D. Batista (2008).
\newblock Cd19 is essential for b cell activation by promoting b cell
  receptor--antigen microcluster formation in response to membrane-bound
  ligand.
\newblock {\em Nature immunology\/}~{\em 9\/}(1), 63--72.

\bibitem[\protect\citeauthoryear{Doz, Giannone, and Reichlin}{Doz
  et~al.}{2012}]{doz2012quasi}
Doz, C., D.~Giannone, and L.~Reichlin (2012).
\newblock A quasi--maximum likelihood approach for large, approximate dynamic
  factor models.
\newblock {\em Review of economics and statistics\/}~{\em 94\/}(4), 1014--1024.

\bibitem[\protect\citeauthoryear{Evers, Sheikhhassani, Haks, Storm, Ottenhoff,
  and Mashaghi}{Evers et~al.}{2022}]{evers2022single}
Evers, T.~M., V.~Sheikhhassani, M.~C. Haks, C.~Storm, T.~H. Ottenhoff, and
  A.~Mashaghi (2022).
\newblock Single-cell analysis reveals chemokine-mediated differential
  regulation of monocyte mechanics.
\newblock {\em Iscience\/}~{\em 25\/}(1).

\bibitem[\protect\citeauthoryear{Hui, Warton, Ormerod, Haapaniemi, and
  Taskinen}{Hui et~al.}{2017}]{hui2017variational}
Hui, F.~K., D.~I. Warton, J.~T. Ormerod, V.~Haapaniemi, and S.~Taskinen (2017).
\newblock Variational approximations for generalized linear latent variable
  models.
\newblock {\em Journal of Computational and Graphical Statistics\/}~{\em
  26\/}(1), 35--43.

\bibitem[\protect\citeauthoryear{Izenman}{Izenman}{1975}]{izenman1975reduced}
Izenman, A.~J. (1975).
\newblock Reduced-rank regression for the multivariate linear model.
\newblock {\em Journal of multivariate analysis\/}~{\em 5\/}(2), 248--264.

\bibitem[\protect\citeauthoryear{Korsunsky, Millard, Fan, Slowikowski, Zhang,
  Wei, Baglaenko, Brenner, Loh, and Raychaudhuri}{Korsunsky
  et~al.}{2019}]{korsunsky2019fast}
Korsunsky, I., N.~Millard, J.~Fan, K.~Slowikowski, F.~Zhang, K.~Wei,
  Y.~Baglaenko, M.~Brenner, P.-r. Loh, and S.~Raychaudhuri (2019).
\newblock Fast, sensitive and accurate integration of single-cell data with
  harmony.
\newblock {\em Nature Methods\/}~{\em 16\/}(12), 1289--1296.

\bibitem[\protect\citeauthoryear{Lakkis, Schroeder, Su, Lee, Bashore, Reilly,
  and Li}{Lakkis et~al.}{2022}]{lakkis2022multi}
Lakkis, J., A.~Schroeder, K.~Su, M.~Y. Lee, A.~C. Bashore, M.~P. Reilly, and
  M.~Li (2022).
\newblock A multi-use deep learning method for cite-seq and single-cell rna-seq
  data integration with cell surface protein prediction and imputation.
\newblock {\em Nature machine intelligence\/}~{\em 4\/}(11), 940--952.

\bibitem[\protect\citeauthoryear{Lee, Chugh, Shen, Eberle, and Dittmer}{Lee
  et~al.}{2013}]{btt091}
Lee, S., P.~E. Chugh, H.~Shen, R.~Eberle, and D.~P. Dittmer (2013, 02).
\newblock {Poisson factor models with applications to non-normalized microRNA
  profiling}.
\newblock {\em Bioinformatics\/}~{\em 29\/}(9), 1105--1111.

\bibitem[\protect\citeauthoryear{Liang and McCullagh}{Liang and
  McCullagh}{1993}]{liang1993case}
Liang, K.-Y. and P.~McCullagh (1993).
\newblock Case studies in binary dispersion.
\newblock {\em Biometrics\/}, 623--630.

\bibitem[\protect\citeauthoryear{Liu, Liao, Luo, Yang, Lau, Jiao, Shi, Zhai,
  Ji, Yeong, et~al.}{Liu et~al.}{2023}]{liu2023probabilistic}
Liu, W., X.~Liao, Z.~Luo, Y.~Yang, M.~C. Lau, Y.~Jiao, X.~Shi, W.~Zhai, H.~Ji,
  J.~Yeong, et~al. (2023).
\newblock Probabilistic embedding, clustering, and alignment for integrating
  spatial transcriptomics data with precast.
\newblock {\em Nature Communications\/}~{\em 14\/}(1), 296.

\bibitem[\protect\citeauthoryear{Liu, Liao, Yang, Lin, Yeong, Zhou, Shi, and
  Liu}{Liu et~al.}{2022}]{liu2022joint}
Liu, W., X.~Liao, Y.~Yang, H.~Lin, J.~Yeong, X.~Zhou, X.~Shi, and J.~Liu
  (2022).
\newblock Joint dimension reduction and clustering analysis of single-cell
  rna-seq and spatial transcriptomics data.
\newblock {\em Nucleic acids research\/}~{\em 50\/}(12), e72--e72.

\bibitem[\protect\citeauthoryear{Liu, Lin, Zheng, and Liu}{Liu
  et~al.}{2023}]{liu2023generalized}
Liu, W., H.~Lin, S.~Zheng, and J.~Liu (2023).
\newblock Generalized factor model for ultra-high dimensional correlated
  variables with mixed types.
\newblock {\em Journal of the American Statistical Association\/}~{\em
  118\/}(542), 1385--1401.

\bibitem[\protect\citeauthoryear{Liu and Zhong}{Liu and
  Zhong}{2024}]{liu2024highdimensional}
Liu, W. and Q.~Zhong (2024).
\newblock High-dimensional covariate-augmented overdispersed poisson factor
  model.
\newblock {\em arXiv preprint arXiv:2402.15071\/}.

\bibitem[\protect\citeauthoryear{Mauri and Bosma}{Mauri and
  Bosma}{2012}]{mauri2012immune}
Mauri, C. and A.~Bosma (2012).
\newblock Immune regulatory function of b cells.
\newblock {\em Annual review of immunology\/}~{\em 30}, 221--241.

\bibitem[\protect\citeauthoryear{McFadden}{McFadden}{1987}]{mcfadden1987regression}
McFadden, D. (1987).
\newblock Regression-based specification tests for the multinomial logit model.
\newblock {\em Journal of econometrics\/}~{\em 34\/}(1-2), 63--82.

\bibitem[\protect\citeauthoryear{McInnes, Healy, Saul, and
  Gro{\ss}berger}{McInnes et~al.}{2018}]{mcinnes2018umap}
McInnes, L., J.~Healy, N.~Saul, and L.~Gro{\ss}berger (2018).
\newblock Umap: Uniform manifold approximation and projection.
\newblock {\em Journal of Open Source Software\/}~{\em 3\/}(29), 861.

\bibitem[\protect\citeauthoryear{Mimitou, Lareau, Chen, Zorzetto-Fernandes,
  Hao, Takeshima, Luo, Huang, Yeung, Papalexi, et~al.}{Mimitou
  et~al.}{2021}]{mimitou2021scalable}
Mimitou, E.~P., C.~A. Lareau, K.~Y. Chen, A.~L. Zorzetto-Fernandes, Y.~Hao,
  Y.~Takeshima, W.~Luo, T.-S. Huang, B.~Z. Yeung, E.~Papalexi, et~al. (2021).
\newblock Scalable, multimodal profiling of chromatin accessibility, gene
  expression and protein levels in single cells.
\newblock {\em Nature biotechnology\/}~{\em 39\/}(10), 1246--1258.

\bibitem[\protect\citeauthoryear{Pang, Zhao, and Wang}{Pang
  et~al.}{2023}]{pang2023factor}
Pang, D., H.~Zhao, and T.~Wang (2023).
\newblock Factor augmented inverse regression and its application to microbiome
  data analysis.
\newblock {\em Journal of the American Statistical Association\/}, 1--11.

\bibitem[\protect\citeauthoryear{She and Chen}{She and
  Chen}{2017}]{she2017robust}
She, Y. and K.~Chen (2017).
\newblock Robust reduced-rank regression.
\newblock {\em Biometrika\/}~{\em 104\/}(3), 633--647.

\bibitem[\protect\citeauthoryear{Stoeckius, Hafemeister, Stephenson,
  Houck-Loomis, Chattopadhyay, Swerdlow, Satija, and Smibert}{Stoeckius
  et~al.}{2017}]{stoeckius2017simultaneous}
Stoeckius, M., C.~Hafemeister, W.~Stephenson, B.~Houck-Loomis, P.~K.
  Chattopadhyay, H.~Swerdlow, R.~Satija, and P.~Smibert (2017).
\newblock Simultaneous epitope and transcriptome measurement in single cells.
\newblock {\em Nature methods\/}~{\em 14\/}(9), 865--868.

\bibitem[\protect\citeauthoryear{Wang and Blei}{Wang and
  Blei}{2019}]{wang2019frequentist}
Wang, Y. and D.~M. Blei (2019).
\newblock Frequentist consistency of variational bayes.
\newblock {\em Journal of the American Statistical Association\/}~{\em
  114\/}(527), 1147--1161.

\bibitem[\protect\citeauthoryear{Westling and McCormick}{Westling and
  McCormick}{2019}]{westling2019beyond}
Westling, T. and T.~McCormick (2019).
\newblock Beyond prediction: A framework for inference with variational
  approximations in mixture models.
\newblock {\em Journal of Computational and Graphical Statistics\/}~{\em
  28\/}(4), 778--789.

\bibitem[\protect\citeauthoryear{Xu, Demmer, and Li}{Xu
  et~al.}{2021}]{xu2021zero}
Xu, T., R.~T. Demmer, and G.~Li (2021).
\newblock Zero-inflated poisson factor model with application to microbiome
  read counts.
\newblock {\em Biometrics\/}~{\em 77\/}(1), 91--101.

\end{thebibliography}

\end{document}